\documentclass[twocolumn]{oja}
\usepackage{lmodern}
\usepackage{amsmath}
\usepackage{xcolor} 
\usepackage{hyperref}
\usepackage{orcidlink}
\usepackage{tabularx}
\usepackage{graphicx}
\usepackage{float}
\usepackage{booktabs}
\usepackage{array}
\usepackage{savesym}
\savesymbol{tablenum} 
\usepackage{siunitx}
\usepackage{threeparttable}
\usepackage{needspace}
\let\oldsection\section
\renewcommand{\section}{%
  \Needspace{4\baselineskip}%
  \oldsection
}
\restoresymbol{SIX}{tablenum} 
\renewcommand\orcidlink[1]{\href{https://orcid.org/#1}{\includegraphics[width=8pt]{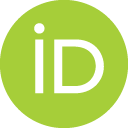}}}

\newcolumntype{C}{>{\centering\arraybackslash}X}
\newcolumntype{L}{>{\raggedright\arraybackslash}X}

\begin{document}

\title{A comprehensive Gaia DR3 study of the open cluster NGC 6134: Bayesian distances, present-day mass function, mass segregation, and structural decomposition}

\shorttitle{Distances, mass function, and structure of NGC 6134}

\author{Agus T. P. Jatmiko$^1$}
\author{Teduh Perhati$^{1,2}$}
\author{Dhimaz G. Ramadhan$^1$}
\author{Muhammad Yusuf$^1$}
\author{Denny Mandey$^1$}
\author{Mochamad I. Arifyanto$^{1,3}$}
\author{Premana W. Premadi$^{1,3}$}
\author{Sahlan Ramadhan$^{4}$}
\author{Ferry Yap$^5$}
\author{Laksma Satya$^5$}

\affiliation{$^1$Bosscha Observatory, FMIPA, Institut Teknologi Bandung, Bandung Barat 40391, Indonesia}
\affiliation{$^2$Master Program in Astronomy, FMIPA, Institut Teknologi Bandung, Bandung 40132, Indonesia}
\affiliation{$^3$Astronomy Research Group, FMIPA, Institut Teknologi Bandung, Bandung 40132, Indonesia}
\affiliation{$^4$Graduate School of Science and Engineering, Ibaraki University, Ibaraki 310-8512, Japan}  
\affiliation{$^5$Undergraduate Program in Astronomy, FMIPA, Institut Teknologi Bandung, Bandung 40132, Indonesia} 
\email{agustriono.pj@itb.ac.id} 

\shortauthors{Jatmiko et al.}

\begin{abstract}
Open clusters are key laboratories for studying the formation and dynamical evolution of stellar populations. We present a comprehensive characterization of the intermediate-age open cluster NGC 6134 using Gaia DR3 astrometry and photometry. Adopting the 890-member Hierarchical Density-Based Spatial Clustering of Applications with Noise (HDBSCAN) membership catalog, we determine the cluster distance and individual member distances with the Bayesian hierarchical code Kalkayotl under a King spatial prior, derive the fundamental parameters with the automated isochrone-fitting code ASteCA, quantify mass segregation, and decompose the projected spatial distribution into structural components. We obtain a Bayesian cluster distance of $1070.8^{+21.0}_{-19.0}$ pc (mode and 68\% credible interval), consistent with the ASteCA value ($d = 1064.4 \pm 15.2$ pc), with member distances peaking at $1071.2 \pm 2.5$ pc. ASteCA yields $\mathrm{[Fe/H]} = 0.08 \pm 0.06$, $\log(t) = 9.14 \pm 0.01$, and $A_\mathrm{V} = 1.03 \pm 0.05$ mag. The present-day mass function (PDMF) follows a broken power law with slopes $\Gamma_A = -2.10^{+0.10}_{-0.10}$ and $\Gamma_B = 1.59^{+0.14}_{-0.14}$ and break mass $M_c = 0.95^{+0.02}_{-0.02}\ M_\odot$, with a negative low-mass slope and a high-mass slope steeper than the canonical Salpeter value ($\Gamma = 1.35$), consistent with the preferential loss of the most massive stars through stellar evolution rather than with dynamical depletion. The observed mass segregation is consistent with an origin in long-term dynamical evolution. The cluster comprises a core, a diffuse halo, and a coherent plane-of-sky elongated feature that is unlikely to arise from noise based on a size-matched isotropic null test, and that lies beyond the radius at which masses are available. The observed main-sequence gap is closely linked to the cluster's binary fraction.
\end{abstract}


\section{Introduction}

Open clusters (OCs) are among the most powerful laboratories for probing the structure and dynamical evolution of the Galactic disk. Because their member stars form nearly simultaneously from a common molecular cloud, OCs provide a rare self-consistent framework in which age, distance, chemical composition, and kinematics are jointly constrained rather than independently fitted. This makes them ideal tracers of spiral-arm structure, the radial metallicity gradient, and the dynamical evolution of the thin disk \citep{Friel:1995, Dias:2002, Cantat-Gaudin:2020}. Two properties in particular set OCs apart from other stellar systems. First, their hierarchical, two-body-relaxed structure is governed by few enough parameters (age, density, velocity dispersion) that internal quantities such as mass segregation can, in principle, be predicted and then tested. Second, their stellar populations are coeval and chemically homogeneous, which removes much of the degeneracy between age and metallicity that complicates the interpretation of field stars.

The Gaia mission \citep{GaiaMission:2016} has made such tests tractable. Gaia Data Release 3 (DR3; \citealt{GaiaDR3:2023}) provides astrometric solutions and three-band photometry ($G$, $G_\mathrm{BP}$, $G_\mathrm{RP}$) for over 1.8 billion sources, enabling membership determinations and structural analyses of unprecedented depth \citep{Bossini:2019, CantatandAnders:2020, Tarricq:2022}. What remains missing is a uniform, self-consistent characterization of individual intermediate-age clusters in which the distance, age, metallicity, internal structure, mass segregation, and mass function are all derived from the same membership catalog and the same reduction pipeline. Without that internal consistency, published values for a single cluster cannot be compared meaningfully, and the physical interpretation of features such as mass segregation or deviations from a universal initial mass function (IMF) remains ambiguous \citep[e.g.][]{Dib:2018,Vesperini:1997}.

NGC 6134 ($\alpha = 246.953^\circ$, $\delta = -49.191^\circ$, J2015.5; \citealp{Cantat-Gaudin:2020}) in the constellation Norma is a useful case in point. Spectroscopic abundance studies report a moderately metal-rich composition of $[\mathrm{Fe/H}] \approx +0.13$ \citep{Carretta:2004, Mikolaitis:2010}, and published ages span approximately $0.7$--$1.4$~Gyr \citep{Mikolaitis:2010, CantatandAnders:2020}. Its color--magnitude diagram is well populated across the main sequence, the turn-off, and the giant branch, which makes the cluster well suited to isochrone fitting, and it lies essentially in the Galactic plane at $\ell_\mathrm{c} \approx 334.9^\circ$, $b_\mathrm{c} \approx -0.2^\circ$. Membership has been determined through both photometric and astrometric means \citep{clariaandMermilliod:1992, Kharchenko:2013, LiuandPang:2019, CantatandAnders:2020, Tarricq:2022, Angelo:2023, Yusuf:2024}, and fundamental parameters have been derived by several groups \citep{Lindoff:1972, Bruntt:1999, PaunzenandMaitzen:2002, Ahumada:2002, Carretta:2004, Mikolaitis:2010, Ahumada:2013}.

The most recent Gaia DR3 study of this cluster, by \citet{Zeng:2025}, derived fundamental parameters through Bayesian isochrone fitting, fitted the radial density profile with a King model, measured the present-day mass function, quantified mass segregation, and identified blue straggler candidates with multi-wavelength follow-up. Three methodological differences motivate the present study. First, the cluster distance is inferred from a distance modulus alone; distances are not determined for individual member stars, so the three-dimensional structure of NGC 6134 remains unconstrained. Second, the binary fraction of the cluster is not quantified, and consequently the physical origin of the main-sequence split noted in this cluster \citep{Ahumada:2013, Antona:2015} is left unaddressed. Third, their candidate selection was confined to a $1\degr$ search radius, which truncates the cluster inside its measured boundary and limits sensitivity to the low-density outer structure examined in Section~\ref{sec:substructures}. Where the two studies overlap, their ages, metallicities, and projected radii agree closely; a detailed comparison, including the distances and reddening, is given in Section~\ref{sec:compare}.

Prior distance determinations for NGC 6134 have relied on a heterogeneous mixture of photometric distance estimates and pre-Gaia astrometric catalogs, producing systematic disagreement between studies \citep{Lindoff:1972, Bruntt:1999, Carretta:2004}. Gaia DR3 individual stellar parallaxes now permit a direct, Bayesian, and internally consistent determination of both the cluster distance and the distances of its individual members, and this forms the starting point of the present work.

Beyond their role as Galactic tracers, OCs are increasingly recognized as useful testbeds for exoplanet formation and evolution, because precise cluster ages and metallicities remove two of the largest uncertainties in the stellar parameters of field stars \citep{Fischer:2005, Meibom:2013}. Dedicated programs such as PATHOS \citep{Nardiello:2021} and the K2 campaigns \citep{Mann:2017, Rizzuto:2018} have demonstrated the value of this approach, but the number of confirmed planets in clusters remains too small to constrain occurrence rates as a function of age or environment. Establishing accurate membership lists and fundamental parameters for southern OCs is therefore a necessary preparatory step. Bosscha Observatory (Lembang, Indonesia) has conducted a long-baseline transit survey of several southern OCs since 2015, using the 35~cm Bosscha Robotic Telescope (BRT; \citealp{Yusuf:2019}) and the 28~cm Survey Telescope for Exoplanet and Variable Stars (STEVia; \citealp{Mandey:2019}) in the Bessell-$R$ band, achieving a photometric precision of $\sim$10~mmag and identifying several periodic variable stars in the target fields.

In this paper we present a self-consistent characterization of NGC 6134 using Gaia DR3 astrometry and photometry. Building on the membership catalog derived by \citet{Yusuf:2024} with the HDBSCAN algorithm \citep{campello:2013} implemented in the \texttt{scikit-learn} Python package \citep{McInnes:2017}, our goals are to: (1) derive a robust cluster distance and the distances of individual member stars from Gaia DR3 parallaxes under an explicit spatial prior; (2) determine the fundamental parameters, including age, metallicity, distance, reddening, and binary fraction, with the automated isochrone-fitting code ASteCA; (3) characterize the internal dynamics by quantifying mass segregation and assessing whether it reflects primordial or dynamical origin; (4) construct the PDMF over the mass range where the primary census is complete; and (5) decompose the projected spatial distribution into structural components and test the coherence of the identified substructures against an isotropic null model. Together these analyses yield a mutually consistent physical picture of NGC 6134 and provide the astrometric and photometric foundation required for the ongoing transit survey at Bosscha Observatory.

This paper is organized as follows. Section \ref{sec:distance} describes the adopted membership catalog and the Bayesian distance determination, including the choice of spatial prior. Section \ref{sec:parameter} presents the determination of the cluster fundamental parameters and binary fraction with ASteCA. Section \ref{sec:substructures} covers the projected and three-dimensional structure of the cluster, the mass-segregation analysis, the PDMF, and the substructure decomposition with its isotropic null test. Section \ref{sec:compare} compares our results with those of previous studies, and Section \ref{sec:conclusion} summarizes our conclusions.

\section{Distance Determination}\label{sec:distance}

\subsection{Membership sample}\label{subsec:membership}
Our previous study \citet{Yusuf:2024} established the membership of NGC 6134 by applying the HDBSCAN algorithm to Gaia DR3 astrometric data and reported 888 members with membership probability $P \geq 0.5$ within 30~pc of the cluster center ($=1.45\degr$ at the reference distance of $1182$~pc \citep{Cantat-Gaudin:2020}). The input query radius was 90~pc ($=4.35\degr$), chosen larger than the expected cluster extent so that the boundary would be measured rather than assumed. Binning the members in 10~pc annuli shows the cumulative count to be complete within 30~pc, and the surface density in the wider cones of Figure 3 of \citet{Yusuf:2024} falls to a low level beyond that radius without any further concentration; the delivered catalog is accordingly truncated at 30~pc, which is why it contains no member beyond $29.8$~pc. We adopt this 30~pc radius, which coincides with the tidal radius to which the King fit of \citet{Yusuf:2024} converges as the cone radius is increased (Figure 3 of \citet{Yusuf:2024}, which displays the 20--70~pc range, the text therein noting that the query extended to 90~pc). We use the membership catalog provided by the authors, which contains 890 entries\footnote{The difference between the 888 members reported in \citet{Yusuf:2024} and the 890 entries in the subsequently supplied catalog may arise from differences in the HDBSCAN realization or computational implementation.}. The determination itself is presented in that paper and is not repeated here; the present study adopts this catalog and exclusively analyzes the member stars it contains. The complete machine-readable membership catalog used throughout this work, including individual membership probabilities, is that of \citet{Yusuf:2024}, deposited at the Centre de Données astronomiques de Strasbourg (CDS) via VizieR under catalog identifier \texttt{J/other/JPhCS/2773.4}\footnote{\url{https://cdsarc.cds.unistra.fr/viz-bin/cat/J/other/JPhCS/2773.4}}.

\subsection{Bayesian approach to distance determination}\label{subsec:bayes_distance}
Gaia DR3 provides parallax measurements for 1.47 billion stars \citep{Vallenari:2023}, but the majority of these sources have low signal-to-noise ratios (SNR $<$ 3). Directly inverting parallax into distance ($r = 1/\varpi$) yields biased and asymmetric estimates whenever the fractional parallax uncertainty is non-negligible \citep{Bailer-Jones:2015, Luri:2018}, and it is formally undefined for negative parallaxes. We therefore avoid any direct parallax-to-distance transformation and instead adopt a forward Bayesian approach in which parallax, not distance, is the observable and the distances are unknowns to be inferred.

While standard Bayesian priors, such as the exponentially decreasing space density (EDSD) model \citep{Astraatmadja:2016a}, are suitable for field stars, cluster members require a dedicated spatial prior. Works such as \citet{Carrera:2019} and \citet{Pang:2022} modified the EDSD prior with a weighted Gaussian component to account for cluster membership probabilities. However, this assumes members are distributed evenly in a simple concentric spherical geometry. A prior that instead encodes the physically motivated density law of a relaxed, tidally truncated cluster is therefore the natural choice for our purpose, and we adopt one below.

For this work, we utilize Kalkayotl \citep{Olivares:2020}, a Bayesian hierarchical framework that performs \emph{forward} (generative) modeling of the astrometric data while accounting for the spatial correlation of Gaia parallax errors \citep{Vasiliev:2019}. The only inputs given to the code are the raw Gaia DR3 observables of the members: each star's parallax $\varpi_i$ and its uncertainty $\sigma_{\varpi,i}$, together with the source coordinates ($\alpha$, $\delta$) needed to build the angular covariance matrix of the correlated parallax errors. A parallax zero-point of $-0.021$~mas \citep{Groenewegen:2021} is subtracted from the observed $\varpi_i$, i.e., $\varpi_i^{\mathrm{corr}} = \varpi_i - (-0.021) = \varpi_i + 0.021$~mas, since Gaia DR3 parallaxes are biased low. Distances are never passed as input; instead, for each star the model draws a `true' distance $d_i$ from the cluster spatial prior, converts it to a predicted parallax $\varpi_i^{\mathrm{pred}} = 1000/d_i$~mas, and assigns a Gaussian likelihood to the observed $\varpi_i$ (with the parallax covariance augmented by the angularly correlated component). The cluster parameters, including the location \texttt{loc}, and the individual distances $d_i$ are therefore \emph{posterior} outputs of the inference rather than inputs. Note that Kalkayotl models the target population as a single physical cluster, requiring the assumption that our selected sample consists entirely of true members (i.e., a 100\% membership probability), neglecting residual field contamination.

Within Kalkayotl, we employ the King prior, which is parameterized by \texttt{loc} (the cluster distance) and \texttt{scl} (the core radius, representing the typical size of the cluster's inner region). We selected the King prior because it is derived directly from stellar dynamics and yields a highly realistic density profile at large radii, naturally accounting for the tidal cutoff radius inherent to gravitationally bound systems. Kalkayotl distributes the distances according to the standardized King formula (i.e., $\mathtt{loc = 0}$ and $\mathtt{scl = 1}$; \citealt{King:1962}, \citealt{Olivares:2020}):

\begin{equation}
	\mathrm{King}(r|r_t) = \frac{\left[ \frac{1}{\sqrt{1+r^2}} - \frac{1}{\sqrt{1+r_t^2}} \right]^2}{2\left[ \frac{r_t}{1+r_t^2} - \frac{2\mathrm{arcsinh}(r_t)}{\sqrt{1+r_t^2}}  + \mathrm{arctan}(r_t) \right]}
\end{equation}
where $r$ is the standardized signed distance offset, supported on $[-r_t,+r_t]$, and $r_t$ is the tidal radius of the cluster, expressed in units of the core radius or core scale (\texttt{scl}) as required by Kalkayotl \citep{Olivares:2020}.

For this model configuration, we applied a parallax zero-point offset of $-0.021$~mas \citep{Groenewegen:2021} and used the cluster distance from \citet{Cantat-Gaudin:2020} as the initial value for the location hyper-parameter. Finally, because our sample features a wide range of parallax uncertainties, we configured the model using the \emph{non-central} (offset) parametrization \citep{Papaspiliopoulos:2007, Betancourt:2015}. The prior itself is identical in both parametrizations; what differs is the geometry in which the sampler explores the joint posterior. In the standard \emph{central} parametrization, each member distance is drawn from the King prior conditional on the current population hyperparameters, so that every $d_i$ is coupled to $\mathtt{loc}$, $\mathtt{scl}$ and $r_t$ through the likelihood. In the non-central parametrization, by contrast, each star draws a \emph{standardized} offset $u_i$ from the unit King distribution, $u_i \sim \mathrm{King}(0, 1, r_t)$, and the physical distance follows deterministically:
\begin{equation}
	d_i = \mathtt{loc} + \mathtt{scl}\,u_i,
\end{equation}
so that, a priori, the $u_i$ are independent of $\mathtt{loc}$ and $\mathtt{scl}$. This removes the strong posterior correlation between the population and per-star parameters (the so-called ``funnel'' geometry that makes Hamiltonian Monte Carlo samplers mix extremely slowly in centered formulations), thereby improving sampling efficiency and ensuring robust convergence in the presence of broad observational errors \citep{Papaspiliopoulos:2007, Betancourt:2015}. The median fractional parallax uncertainty of the member sample, which sets the width of the per-star likelihood, is quoted below in connection with the line-of-sight geometry.

\begin{figure*}[t!]
	\centering
	\includegraphics[width=0.7\textwidth]{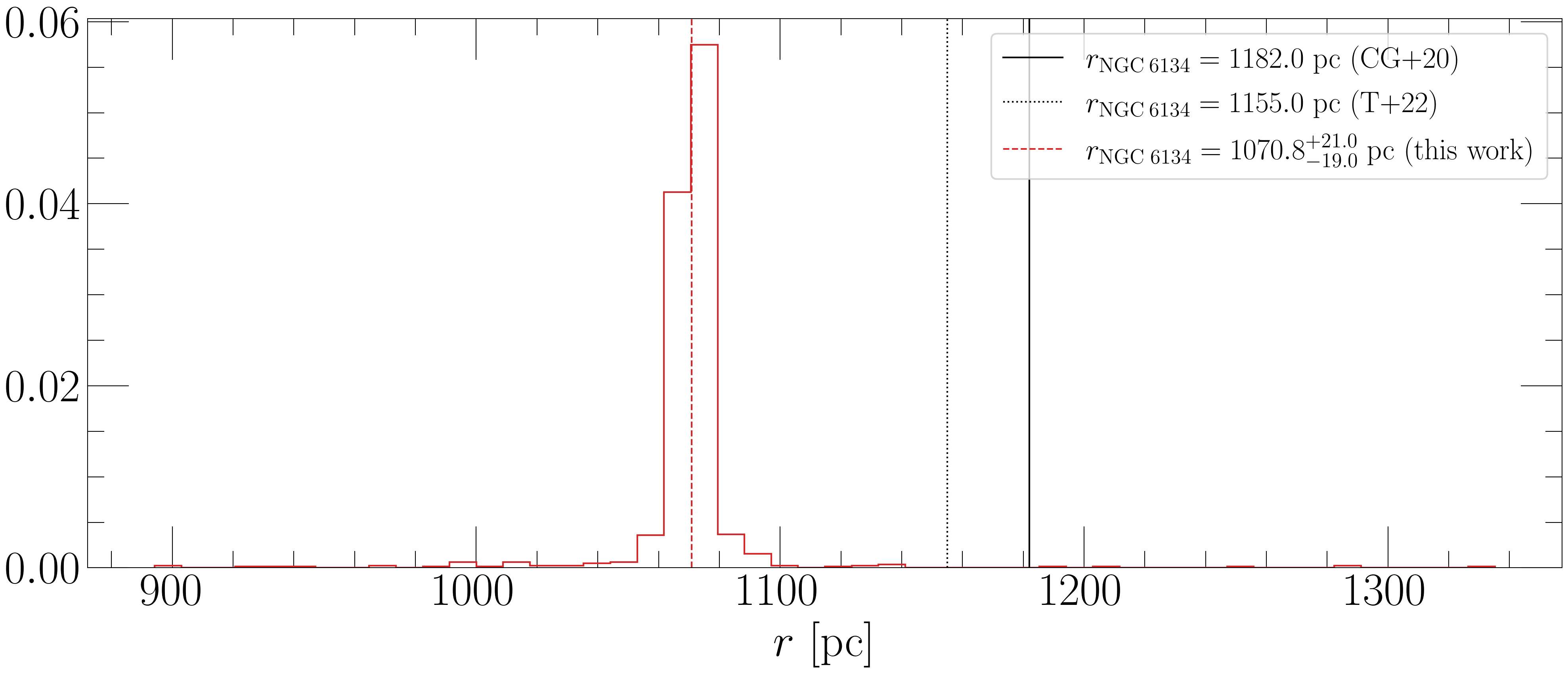}
	\caption{Distance estimation distribution of NGC 6134 star members using Bayesian inference with King prior. The cluster distance (mode of the King location parameter; $1070.8$~pc) from this work is shown. The comparison with the mean distances from \citet{Cantat-Gaudin:2020} and \citet{Tarricq:2022} is discussed in Section \ref{sec:compare}.}
	\label{fig:dist-estimation}
\end{figure*}

The resulting distance distribution for the NGC 6134 members estimated by Kalkayotl is shown in Figure \ref{fig:dist-estimation}. From the posterior we obtain a cluster distance of $\mathrm{loc} = 1070.8$~pc (mode; 68\% credible interval $1051.8$--$1091.8$~pc), a core scale $\mathrm{scl} = 10.9$~pc, and a tidal parameter $r_t = 45.5$ (in units of scl), corresponding to a line-of-sight truncation of $\mathrm{scl}\,r_t \approx 495$~pc. We emphasize that these parameters describe the one-dimensional \emph{line-of-sight} distance distribution within the hierarchical prior and are therefore not directly comparable with projected (2D) King fits to the radial density profile, which we examine in Section \ref{sec:substructures}.

The quoted credible interval requires a caveat. Combining the median fractional parallax uncertainty of the member sample with the sample size, the parallaxes alone would constrain the cluster distance to the level of a few parsecs, an order of magnitude tighter than the interval reported above. The posterior is therefore dominated by the adopted line-of-sight spatial prior rather than by the astrometry: because that prior permits individual members to lie at substantial offsets along the line of sight, offsets the data cannot exclude, the inferred core scale $\mathrm{scl}$ and tidal parameter $r_t$ reflect the functional form of the prior and are not independently measured quantities. The cluster \emph{distance}, by contrast, is robust: it is consistent with the independent photometric determination of Section \ref{sec:parameter} to within its uncertainties, and the ASteCA and astrometric results agree despite resting on different data and methods.

The projected structure of the cluster and the geometry of its members in three dimensions are examined in Section \ref{sec:substructures}.

\section{Parameters Determination}\label{sec:parameter}
\subsection{Automated stellar cluster analysis (ASteCA)}\label{subsec:asteca-params}
Deriving cluster parameters by overlaying theoretical isochrones on the observed color--magnitude diagram (CMD) is subjective, scales poorly, and yields no robust statistical uncertainties \citep[e.g.][]{Kharchenko:2013}. Synthetic CMD fitting addresses this by simulating a full two-dimensional stellar population, including the initial mass function, binarity, and photometric errors, and comparing it statistically to the data \citep[e.g.][]{Girardi:2009}. Artificial neural network (ANN) approaches offer an alternative automated route \citep{Cantat-Gaudin:2020}. We adopt ASteCA, which performs this synthetic CMD comparison internally through parallel tempering.

We used ASteCA v0.4.3 \citep{Perren:2015} to determine the cluster age $\log(t)$, metallicity $\mathrm{[Fe/H)}$, distance modulus $(m-M)_0$, and visual absorption $A_\mathrm{V}$. In brief, ASteCA builds a synthetic isochrone from a theoretical stellar evolution model at fixed metallicity and age, displaces it in distance modulus and extinction until it best matches the observed CMD, and infers the posterior distribution of the parameters from that match. ASteCA has been applied to a number of clusters \citep[e.g.][]{Scholz:2015, Ghosh:2022, Rain:2024}.

ASteCA additionally reports the binary fraction $f_\mathrm{bin}$, the fraction of detected member systems that are binaries, $f_\mathrm{bin}\equiv N_\mathrm{bin}/(N_\mathrm{bin}+N_\mathrm{single})$, where unresolved (near-)equal-mass pairs appear in the CMD as systems lying above the single-star main sequence. This quantity is not one of the fitted parameters: the fit solves for $(m-M)_0$, $z$, $\log(t)$, and $A_\mathrm{V}$ only, each under a flat bounded prior, with no prior imposed on $f_\mathrm{bin}$. It is instead inferred from the binary population of the synthetic clusters. Each synthetic star is drawn as a binary with a mass-dependent probability $p_\mathrm{b}(m_1)=\alpha+\beta\,\log_{10}(m_1/M_\odot)$, so that the binary fraction of a synthetic population is the corresponding average of $p_\mathrm{b}$ over its members; ASteCA returns this quantity for the population of accepted models, and we quote that output. It is therefore a property of the imposed binary model rather than an independent measurement of the individual binary population.

ASteCA reads the membership probabilities from our HDBSCAN catalog ($P \geq 0.5$) and applies its cell-based photometric cleaning to the color--magnitude diagram, restricting to its fitted cluster region and then discarding, cell by cell, the lowest-probability stars in proportion to the local field density. This retains 655 clean members for the isochrone fitting, which is performed in four parameter spaces that include the distance modulus, $z$, $\log(t)$, and $A_\mathrm{V}$, over the defined parameter range using the parallel tempering algorithm. The result is the probability distribution of each parameter which includes its mean and standard deviation.

We construct the underlying synthetic isochrones from the PARSEC isochrone grid version 1.2S \citep{Bressan:2012} taken with CMD v3.6\footnote{\url{http://stev.oapd.inaf.it/cgi-bin/cmd}}, spanning $\log(t) = 7.00$ to $10.10$ in steps of $\Delta \log(t) = 0.05$ and $z = 0.0005$ to $0.0305$ in steps of $\Delta z = 0.0100$, over which $z$ and $\log(t)$ are interpolated. The distance modulus and $A_\mathrm{V}$ are instead sampled as continuous variables from flat, bounded priors over the ranges $(m-M)_0 = 10$--$14$ mag and $A_\mathrm{V} = 0$--$4$ mag, and applied to the isochrone as continuous shifts; they are therefore not discretized by any grid step. The fitted $(m-M)_0 = 10.136$~mag lies inside its prior, $0.14$~mag from the nearest bound, so the posterior is not materially truncated by it. We also used the IMF of \citet{Kroupa:2002} and the Gaia passband transmission curves, for which the EDR3-calibrated response functions are those distributed with Gaia and applied to DR3 photometry. As input, we used Gaia DR3 photometry data for the cluster members, as described in Section \ref{subsec:membership}, and adopted the cluster center of NGC 6134 from \citet{Cantat-Gaudin:2020}. Additionally, we set the parameter fitting duration to 200 minutes while keeping the remaining parameters at their default values in the ASteCA configuration file.

In ASteCA, binarity is built into the synthetic-population generator through a primary-mass-dependent binary probability in logarithmic form, $p_\mathrm{b}(m_1)=\alpha+\beta\,\log_{10}(m_1/M_\odot)$, together with a power-law mass-ratio distribution for the companions \citep{Perren:2015}. Following the multiplicity fraction of field stars \citep{Duchene:2013}, we held the binarity coefficients fixed at $\alpha=0.3$ and $\beta=0.15$; both are inputs rather than fitted quantities. The mass-ratio distribution is the mass-dependent law of \citet{Duchene:2013}, also fixed. Unresolved binaries populate each synthetic CMD as systems scattered above the single-star isochrone (by up to $\sim0.75$ mag in $G$ for near-equal-mass pairs), and the observed excess of such systems is what constrains the binary population relative to the observed CMD. Because the coefficients governing this synthetic binary population are fixed inputs rather than fitted quantities, the recovered $f_\mathrm{bin}$ is constrained by, and will tend toward, the imposed binary probability law; we therefore regard $f_\mathrm{bin}$ as a measurement consistent with, rather than independent of, the adopted $\alpha$ and $\beta$, and report the $\pm 0.02$ as the reproducibility of the estimator rather than a precision on the underlying binary fraction of the cluster.

We ran ASteCA 20 times to obtain the cluster parameters, including $\log(t)$, metallicity in $z$, $(m-M)_0$, $A_{\mathrm{V}}$, and $f_\mathrm{bin}$. To convert metallicity to $\mathrm{[Fe/H]}$, we used the relation $\mathrm{[Fe/H]} = \log(z/z_{\odot})$ with $z_{\odot} = 0.0152$ \citep{Bressan:2012}. The distance modulus was converted to distance $d$ using $(m-M)_0 = -5 + 5\log(d)$. Each parameter is combined across the realizations by inverse-variance weighting, $\bar{x} = \left(\sum_i w_i x_i\right)/\left(\sum_i w_i\right)$ with $w_i = 1/\sigma_i^2$, in which the $\sigma_i$ entering the weights is the posterior standard deviation reported by run $i$ for that parameter, not the scatter between runs. The quoted uncertainty is a separate quantity: the run-to-run standard deviation of the run values, which measures the reproducibility of the sampler under repeated fits of the same data. We adopt this latter quantity throughout, as it is the larger of the two for every parameter except the distance modulus and metallicity, where the within-run posterior width is instead the more conservative measure (see below). Because $\mathrm{[Fe/H]}$ and $d$ are strictly monotonic but nonlinear functions of the fitted variables, $z$ and $(m-M)_0$, the $\pm1\sigma$ bounds are propagated through the exact transformation itself, $[\mathrm{Fe/H}]_{\pm} = \log\!\left[(z\pm\sigma_z)/z_{\odot}\right]$ and $d_{\pm} = 10^{(\mu\pm\sigma_\mu+5)/5}$, which yields asymmetric errors on $\mathrm{[Fe/H}]$ and $d$; the asymmetry is negligible for $d$ (well below the reported rounding), so its error is quoted symmetrically. The results are as follows:

\begin{enumerate}
	\setlength{\itemsep}{0pt}
	\item[(i)] $\mathrm{[Fe/H]} = 0.08^{+0.056}_{-0.065}$
	\item[(ii)] $\log(t) = 9.14 \pm 0.01$
	\item[(iii)] $A_\mathrm{V} = 1.03 \pm 0.05$
	\item[(iv)] $(m-M)_0 = 10.136 \pm 0.031$~mag
	\item[(v)] $d\; \mathrm{(pc)} = 1064.4 \pm 15.2$
	\item[(vi)] $f_\mathrm{bin} = 0.42 \pm 0.02$
\end{enumerate}

\begin{figure}[t!]
	\centering
	\includegraphics[width=0.48\textwidth]{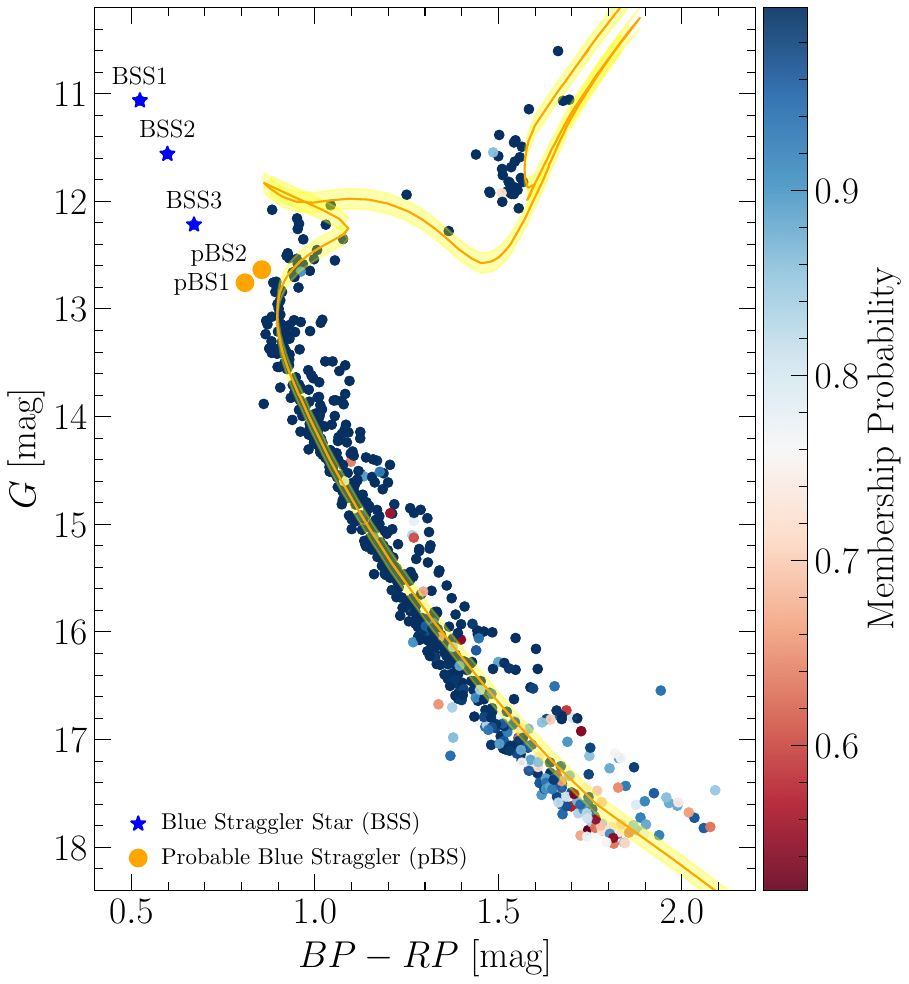}
	\caption{Color--magnitude diagram of NGC 6134. The color scale indicates the membership probability of each star. The solid black line represents the best-fit PARSEC isochrone, with the shaded area indicating its $3\sigma$ confidence region. Confirmed blue straggler stars (BSSs) are marked by blue star symbols, and blue straggler candidates (pBS) are denoted by filled orange circles. Details for the individually labeled stars are provided in Table \ref{tab:bss}.} 
	\label{fig:isochrone}
\end{figure}

Two caveats apply to these uncertainties. First, the quoted errors are run-to-run scatter, whereas the width of the within-run posterior is somewhat larger for two of the parameters. For the distance modulus the within-run posterior width is $0.048$~mag, corresponding to approximately $24$~pc; the run-to-run scatter of $\pm 15.2$~pc quoted above is therefore the more optimistic of the two available estimates, and we regard the distance uncertainty as being of order $15$--$24$~pc. For metallicity the within-run posterior is also broader than the run-to-run figure, spanning approximately $0.09$~dex against the quoted $0.06$~dex. This reflects the deliberately coarse metallicity grid, which comprises only four nodes over $z = 0.0005$--$0.0305$; although $z$ and $\log(t)$ are sampled as continuous parameters and interpolated between nodes, the attainable precision in $[\mathrm{Fe/H}]$ is limited by that interpolation rather than by the statistics of the fit. We accordingly regard the quoted metallicity uncertainty as a lower bound on the true uncertainty in this parameter. All other quoted values, $\log(t)$ and $A_\mathrm{V}$, have within-run posteriors tighter than the run-to-run scatter and are therefore unaffected.
The distance estimates derived using ASteCA with photometric data and those obtained in Section \ref{sec:distance} from astrometric data are consistent within their respective uncertainties. To display the results of the isochrone fitting, we query the isochrone data with the weighted average of [Fe/H] and $\log(t)$. Then we shift the queried isochrone on the vertical axis of the CMD by $(m-M)_0 + A_\mathrm{G}$, with $A_{\mathrm{G}} = 0.789 A_{\mathrm{V}}$, and on the horizontal axis to $E(G_\mathrm{BP}-G_\mathrm{RP})$, with $E(G_\mathrm{BP}-G_\mathrm{RP}) = A_\mathrm{G}/1.89$ \citep{wang:2019}. The results of the isochrone fitting in the CMD, including the $3\sigma$ confidence region, are shown in Figure \ref{fig:isochrone}.

\section{Structure, Mass Segregation, and Substructures}\label{sec:substructures}

\subsection{Projected structure and line-of-sight geometry}\label{subsec:structure}

\begin{figure*}[t!]
	\centering
	\includegraphics[width=0.5\textwidth]{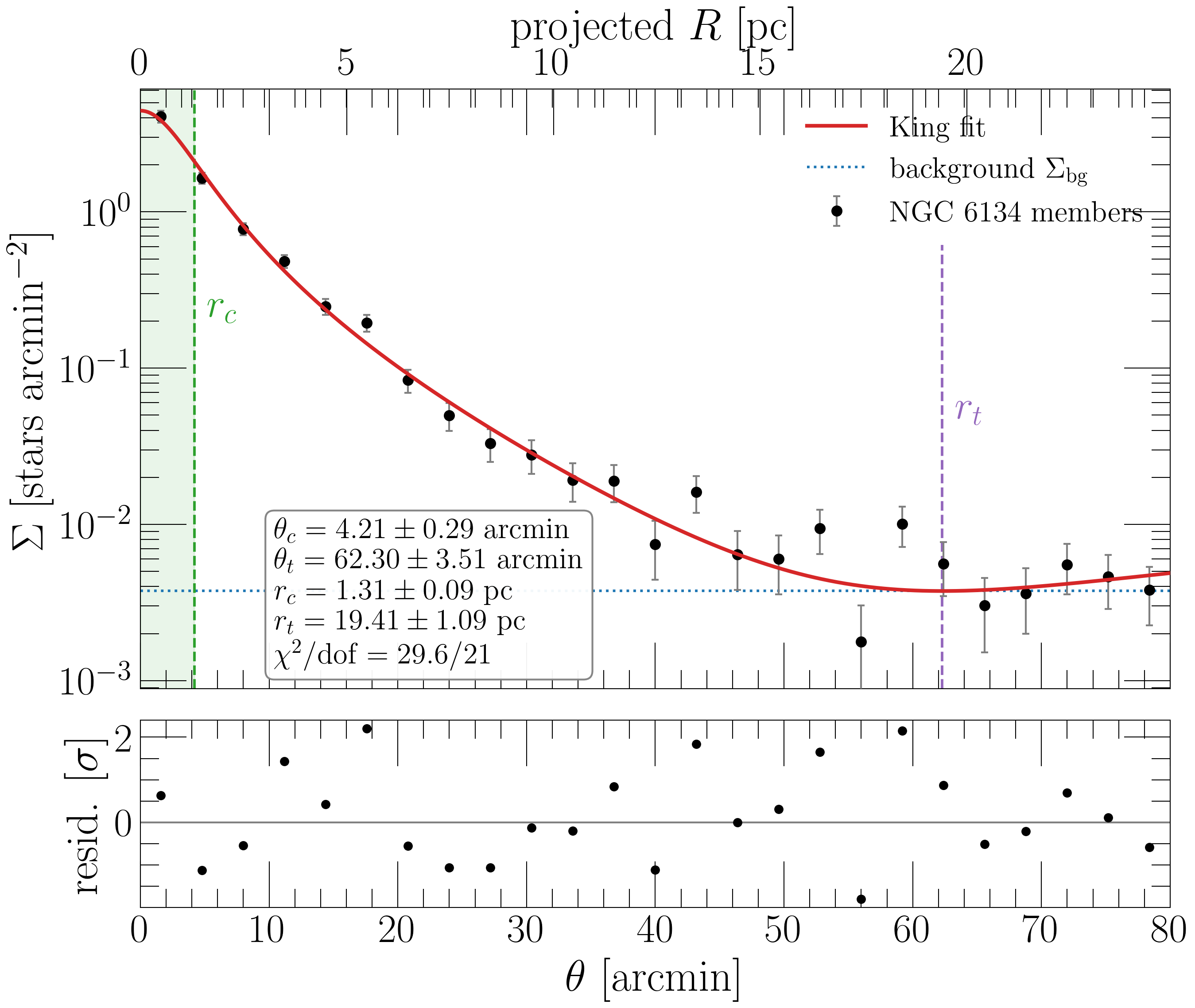}
	\caption{Azimuthally averaged projected radial density profile (RDP) of the 890 NGC 6134 members (black points with Poisson error bars) and the best-fitting projected King model (red curve; dotted line: constant background level; dashed lines: fitted core and tidal radii). The upper panel uses a logarithmic surface-density scale so that the tidal wings remain visible against the core; the lower panel is linear and shows the fit residuals, discussed in the text. The profile is constructed in angular separation from the fitted cluster center so that its shape assumes no distance; the top axis shows the corresponding physical radius obtained by adopting the Kalkayotl distance of $1070.8$~pc.}
	\label{fig:king-rdp}
\end{figure*}

To characterize the projected structure independently, we fit a King surface-density profile to the azimuthally averaged radial density profile (RDP) of the full 890-member catalog. Because the RDP is a purely geometric and magnitude-independent measurement, we do not apply the photometric decontamination used for the isochrone fitting in Section \ref{sec:parameter}; the full membership catalog is the appropriate sample for this profile. To keep the fit independent of any adopted distance, the profile is constructed in angular separation from the fitted cluster center and fitted in angular units, with the distance entering only in the final conversion to parsecs. The fit (Figure~\ref{fig:king-rdp}) yields a core radius of $\theta_c = 4.21 \pm 0.29$~arcmin ($r_c = 1.31 \pm 0.09$~pc) and a tidal radius of $\theta_t = 62.3 \pm 3.5$~arcmin ($r_t = 19.4 \pm 1.1$~pc).

This fit establishes that the cluster has a centrally concentrated profile with a well-defined tidal truncation \emph{in projection}, which makes a centrally concentrated line-of-sight prior physically reasonable. It does not, however, test the line-of-sight profile itself: the two constraints probe orthogonal projections of the same three-dimensional structure. The distinction is quantified by the numbers themselves, since the projected tidal radius of $19.4$~pc and the $495$~pc line-of-sight truncation implied by the Kalkayotl prior differ by more than an order of magnitude. The latter is a property of the adopted prior rather than a measured quantity, as discussed in Section \ref{sec:distance}. Because this fit is restricted to the catalogued members rather than the surrounding field, its tidal radius is not directly comparable with the larger value near $30$~pc obtained by \citet{Yusuf:2024} from progressively wider search cones: those wider cones admit an increasing field contribution, which a King profile absorbs by extending its truncation, whereas the present fit measures the profile of confirmed members with an explicit constant background term.

The fit yields $\chi^2/\mathrm{dof} = 29.6/21 \approx 1.4$. This modest excess is not distributed uniformly across the profile. The six innermost annuli, lying within $19.2$~arcmin ($4.6\,\theta_c$) and containing 665 of the 890 members, contribute $\sim$31\% of the total $\chi^2$, while the nineteen outer annuli contribute the remaining $\sim$69\%. The outer contribution is driven by the least populated bins, where Poisson counting statistics dominate; the residuals there change sign frequently rather than forming a coherent pattern, as expected for counting noise. The excess therefore reflects stochastic fluctuation in the outskirts rather than a systematic deviation from the King form, and the well-sampled core, which is what constrains $\theta_c$, is fitted adequately. We nevertheless note that the fit quality degrades in the outskirts, and the fitted $\theta_t = 62.3 \pm 3.5$~arcmin should be read as an approximate truncation radius rather than a sharply determined quantity. This is consistent with the asymmetric tidal structure revealed by the mixture-model decomposition in Section \ref{subsec:gmm}, which a smooth spherically symmetric profile cannot reproduce.

The projected profile above constrains only the plane of sky. To examine the line-of-sight geometry as well, we combine the angular separations with the per-member distance posterior modes of Section \ref{sec:distance}, and place the 890 members in a heliocentric Cartesian frame using the standard transformation \citep{binney1998galactic}:
\begin{equation}
	\begin{aligned}
		X &= r \cos b \cos \ell \\
		Y &= r \cos b \sin \ell \\
		Z &= r \sin b
	\end{aligned}
	\label{eq:galcart}
\end{equation}
where \( r \) represents the distance of an individual star, and \( \ell \) and \( b \) denote the galactic longitude and latitude of the cluster member, respectively. The Galactic coordinates were obtained by transforming each member's Gaia DR3 ICRS equatorial coordinates \( (\alpha,\delta) \) to the Galactic frame using the standard ICRS-to-Galactic rotation (epoch of J2000; the cluster subtends only \(\sim\)2.7\degr{} on the sky). The 890 members span \(\ell \in [333.62\degr, 336.30\degr]\) and \(b \in [-1.60\degr, +1.10\degr]\). The adopted cluster center, obtained from Gaussian fits to the member \(\ell\) and \(b\) distributions, is \(\ell_\mathrm{c} = 334.91 \pm 0.15\degr\) and \(b_\mathrm{c} = -0.21 \pm 0.13\degr\); NGC 6134 therefore lies essentially in the Galactic plane \((|b| \sim 0.2\degr)\). At this center the heliocentric Cartesian position evaluates to \(X \approx 970\)~pc, \(Y \approx -454\)~pc, \(Z \approx -4\)~pc.

\begin{figure*}[t!]
	\centering
	\includegraphics[width=0.75\textwidth]{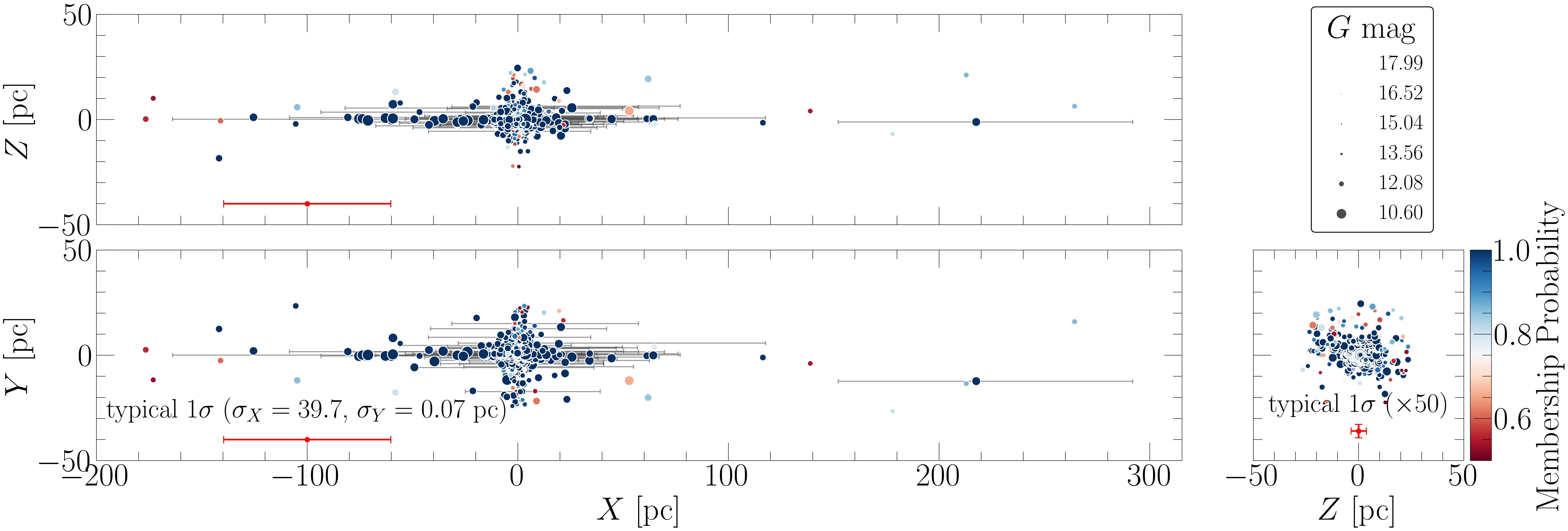}
	\caption{The 3D spatial distribution of the 890 members in the local sky-offset (tangent-plane) frame centered on the cluster \((\ell_\mathrm{c}, b_\mathrm{c}) = (334.91\degr, -0.21\degr)\), projected into the XY, XZ, and YZ planes. The $X$ axis runs along the line of sight (heliocentric distance relative to the cluster distance of $1070.8$~pc), $Y$ along increasing $\ell$, and $Z$ along increasing $b$; the cluster is essentially in the Galactic plane, so its $Z$ extent is small. The marker size represents the \( G \)-band magnitude from Gaia DR3, with larger markers indicating brighter stars, while membership probabilities are color-coded according to the color bar. The cluster appears elongated along $X$ due to distance uncertainties of individual member stars, which smear their distribution along the line of sight. Overplotted is the typical $1\sigma$ (68\% credible interval) per-member uncertainty derived from the Kalkayotl distance posteriors: $\sigma_X = 39.7$~pc (line of sight) compared with $\sigma_Y = 0.07$~pc and $\sigma_Z = 0.06$~pc (tangential). In the $XY$ and $XZ$ panels the red cross near the lower left shows the typical error at native scale, dominated by the horizontal (line-of-sight) bar; in the $YZ$ panel the tangential typical error is magnified by a factor of 50 (lower center, labeled $\times50$). The thin black error bars in the $XY$ and $XZ$ panels trace the asymmetric line-of-sight uncertainty ($\mathrm{mode}-\mathrm{lower}$, $\mathrm{upper}-\mathrm{mode}$) of a random subsample of 45 members.}
	\label{fig:3d-spatial-distribution}
\end{figure*}

For visualization we re-center the distribution on the cluster and express it in the local sky-offset (tangent-plane) frame, in which the \(X\) axis runs along the line of sight (heliocentric distance relative to the cluster distance), \(Y\) along increasing \(\ell\), and \(Z\) along increasing \(b\). The 3D spatial distribution projected into the XY, XZ, and YZ planes is shown in Figure \ref{fig:3d-spatial-distribution}. Note that the \(X\) axis in this frame is \emph{not} the \(X\) axis of Eq.~(\ref{eq:galcart}), which is heliocentric; within the \(\sim\)2.7\degr{} angular footprint of the cluster the two frames are numerically equivalent up to an overall rotation, but they should not be compared axis by axis.

The posterior distance modes of the individual members peak at a median of $1071.2$~pc with a robust scatter of $\pm 2.5$~pc (median absolute deviation, MAD), a dispersion that reflects the weak line-of-sight constraint discussed in Section \ref{sec:distance} rather than a direct measurement of the cluster's line-of-sight thickness; the median was specifically chosen to mitigate the inherent skewness of distance estimates and to minimize the influence of spatial outliers. In 3D space, the cluster consequently appears strongly elongated along the observational axis. This morphological stretching is a well-known artifact driven by inherent uncertainties in individual stellar distance estimates, which preferentially smear the spatial projection along the line of sight. To quantify this effect, we propagate the per-member Kalkayotl distance posterior (mode and 68\% credible interval) into the sky-offset frame: the line-of-sight term carries a typical uncertainty of $\sigma_X \approx 40$~pc (median $39.7$~pc; $16^{\mathrm{th}}$--$84^{\mathrm{th}}$ percentiles $30.8$--$50.7$~pc). This value follows directly from the parallax errors: the median fractional parallax error $f \equiv \sigma_\varpi/\varpi$ of the 890 members is $f \approx 0.044$ ($16^{\mathrm{th}}$--$84^{\mathrm{th}}$ percentiles $0.021$--$0.101$), which at the median parallax of $0.91$~mas corresponds to a fractional distance uncertainty of the same order, and hence to $\sigma_X$ of several tens of parsecs. Because a member's tangential offset scales with its angular separation from the cluster center, the same distance uncertainty contributes only $\sigma_Y \approx 0.07$~pc and $\sigma_Z \approx 0.06$~pc (median; $84^{\mathrm{th}}$ percentile $\lesssim 0.3$~pc) in the tangential directions, while the Gaia DR3 positional errors contribute negligibly. This more than two orders-of-magnitude anisotropy is shown in Figure~\ref{fig:3d-spatial-distribution}: the elongation along $X$ is an artifact of the line-of-sight distance uncertainties, whereas the substructure in the tangential plane is genuine at the level of several parsecs.

\subsection{Mass Segregation}\label{subsec:mass-segregation}

To quantify the degree of mass segregation within the cluster, we employ the mass segregation ratio ($\Lambda_{\text{MSR}}$) formulation of the Minimum Spanning Tree (MST) method introduced by \citet{Allison:2009}. The fundamental premise of this method is to compare the spatial concentration of the $N$ most massive stars against the baseline spatial distribution of $N$ randomly selected cluster members.

Our implementation introduces critical refinements to the standard $\Lambda_{\text{MSR}}$ calculation by utilizing exact spherical geometry, accounting for unresolved binary systems probabilistically, and mitigating known small-number statistical biases.

\subsubsection{Spherical Geometry and Exact Angular Distances}
Standard implementations of the MST method often rely on projecting equatorial coordinates onto a flat 2D Cartesian tangent plane using a gnomonic projection. However, the gnomonic projection is neither conformal nor equidistant. The projection introduces a systematic, non-isotropic Euclidean distortion that scales with the square of the tangent angle, reaching a maximum radial scale factor distortion of $\approx 0.12\%$ at $2^\circ$ and $\approx 0.27\%$ at $3^\circ$ from the field center. While seemingly minor, Kruskal's algorithm is highly sensitive to the strict rank-ordering of pairwise distances. Empirical testing over a $3^\circ$ cluster field demonstrates that this localized geometric stretching is sufficient to alter the fundamental MST edge topology in approximately $28.5\%$ of random spatial realizations. To eliminate this topological instability, our methodology abandons the 2D projection entirely and computes pairwise angular separations exactly on the celestial sphere using the haversine formula \citep{Olivares:2018}.

\subsubsection{Stochastic System Mass Determination}
Rather than relying on static, single-value mass estimates, stellar masses are derived probabilistically to strictly account for measurement uncertainties and the presence of unresolved binaries. We utilize a Monte Carlo framework consisting of 100 iterations. In each iteration, the dynamical system mass for each star is drawn using the following protocol:
\begin{itemize}
  \item Primary Mass: The primary mass ($m_1$) is sampled from a Gaussian distribution, $\mathcal{N}(\mu_{m1}, \sigma_{m1})$, and bounded below by a physical hydrogen-burning floor of $0.08 \, M_\odot$.
  \item Binary Designation: The binary status of the system is evaluated as a Bernoulli trial, drawn against the star's calculated binary probability ($p_{\text{bin}}$).
  \item Secondary Mass: If designated as a binary system, the secondary mass ($m_2$) is conditionally sampled from $\mathcal{N}(\mu_{m2}, \sigma_{m2})$ and clipped to a minimum of $0.0 \, M_\odot$ to prevent the unphysical addition of sub-zero masses.
  \item Total System Mass: The dynamical mass is taken as $m_1 + m_2$ (if binary) or $m_1$ (if single).   
\end{itemize}
The cluster members are subsequently re-sorted in descending order for each iteration based on these dynamic, probabilistically drawn system masses. The Gaussian parameters ($\mu_{m1}, \sigma_{m1}$) and ($\mu_{m2}, \sigma_{m2}$) and the binary probability $p_{\text{bin}}$ entering this protocol are not free or assumed quantities; each is a per-star output of ASteCA's mass-assignment post-processing (\citealp{Perren:2015}, Section~\ref{subsec:asteca-params}). For every accepted isochrone model, ASteCA generates a synthetic cluster that reproduces the photometric errors and binary population of the observed CMD and, using a KD-tree nearest-neighbour search in the ($G$, $G_\mathrm{BP}-G_\mathrm{RP}$) plane, assigns each observed member the mass of its closest synthetic counterpart. Averaging these assignments over the accepted models yields, for each member, the mean primary mass $\mu_{m1}$ and its standard deviation $\sigma_{m1}$ (so that $\sigma_{m1}$ propagates both the photometric scatter and the fitted-parameter uncertainty into $m_1$); likewise, the mean and dispersion of the secondary masses of the binary synthetic counterparts define $\mu_{m2}$ and $\sigma_{m2}$. The binary probability $p_{\text{bin}}$ is ASteCA's per-member Bayesian probability that the photometry is better described by an unresolved binary pair than by a single star, $p_{\text{bin}} = \frac{1}{N_\mathrm{mod}}\sum_k \left[1 + \mathcal{L}_{k,\mathrm{single}}/\mathcal{L}_{k,\mathrm{binary}}\right]^{-1}$, with $\mathcal{L}$ the nearest-neighbour likelihoods of the $N_\mathrm{mod}$ accepted models \citep{Perren:2015}. This per-member quantity drives the binary draw in the Monte Carlo; it is distinct from the aggregate $f_\mathrm{bin}$ reported in Section~\ref{subsec:asteca-params}, which is a property of the imposed binary model rather than a mean of these per-member probabilities.

A secondary mass is drawn only for members that are (i) drawn as binaries and (ii) carry a valid ASteCA companion estimate ($\mu_{m2} > 0$), and any negative draw is clipped at $0\,M_\odot$.

\subsubsection{Robust $\Lambda_{\text{MSR}}$ Formulation}
For a given subset of the $N$ most massive systems, the MST path length ($l_{\text{massive}}$) is calculated from the haversine separations of that subset. The mass segregation ratio is formulated as:
\begin{equation}
  \Lambda_{\text{MSR}} = \frac{\text{Median}(l_{\text{random}})}{l_{\text{massive}}}  
\end{equation}

We deliberately deviate from the original \citet{Allison:2009} formulation by utilizing the median of 500 random background lengths rather than the arithmetic mean. As demonstrated by structural critiques (e.g., \citealp{Parker:2015}), the arithmetic mean is highly susceptible to positive skew from rare outlier draws containing unphysically long edges, which artificially suppresses the measured $\Lambda_{\text{MSR}}$ signal. The median provides a robust normalizer. A ratio $\Lambda_{\text{MSR}} > 1$ indicates mass segregation, while $\Lambda_{\text{MSR}} \approx 1$ indicates a spatial distribution consistent with the random cluster background.

We also track and isolate two distinct sources of uncertainty: (i) Random Sampling Uncertainty: The 16$^\mathrm{th}$ and 84$^\mathrm{th}$ percentiles of the spatial permutations derived from the pre-computed random background. This represents the inherent geometric variance of the cluster itself, and (ii) Mass Propagation Spread: The $1\sigma$ variance in the final $\Lambda_{\text{MSR}}$ values across the Monte Carlo mass iterations. Because stellar mass rankings shift stochastically due to binary additions and Gaussian measurement errors, this band represents the true impact of photometric uncertainty on the dynamical interpretation.

Finally, in adherence to the topological warnings established by \citet{Parker:2015}, we explicitly isolate the regime where $N_{\text{MST}} < 20$. In this small-$N$ regime, stochastic geometric clustering can produce artificially massive $\Lambda_{\text{MSR}}$ peaks that mimic segregation but do not reflect global two-body relaxation. We therefore confine our interpretation of the cluster's dynamical state to the subset where $N_{\text{MST}} \geq 20$.

\begin{figure*}[t!]
	\centering
	\includegraphics[width=0.6\textwidth]{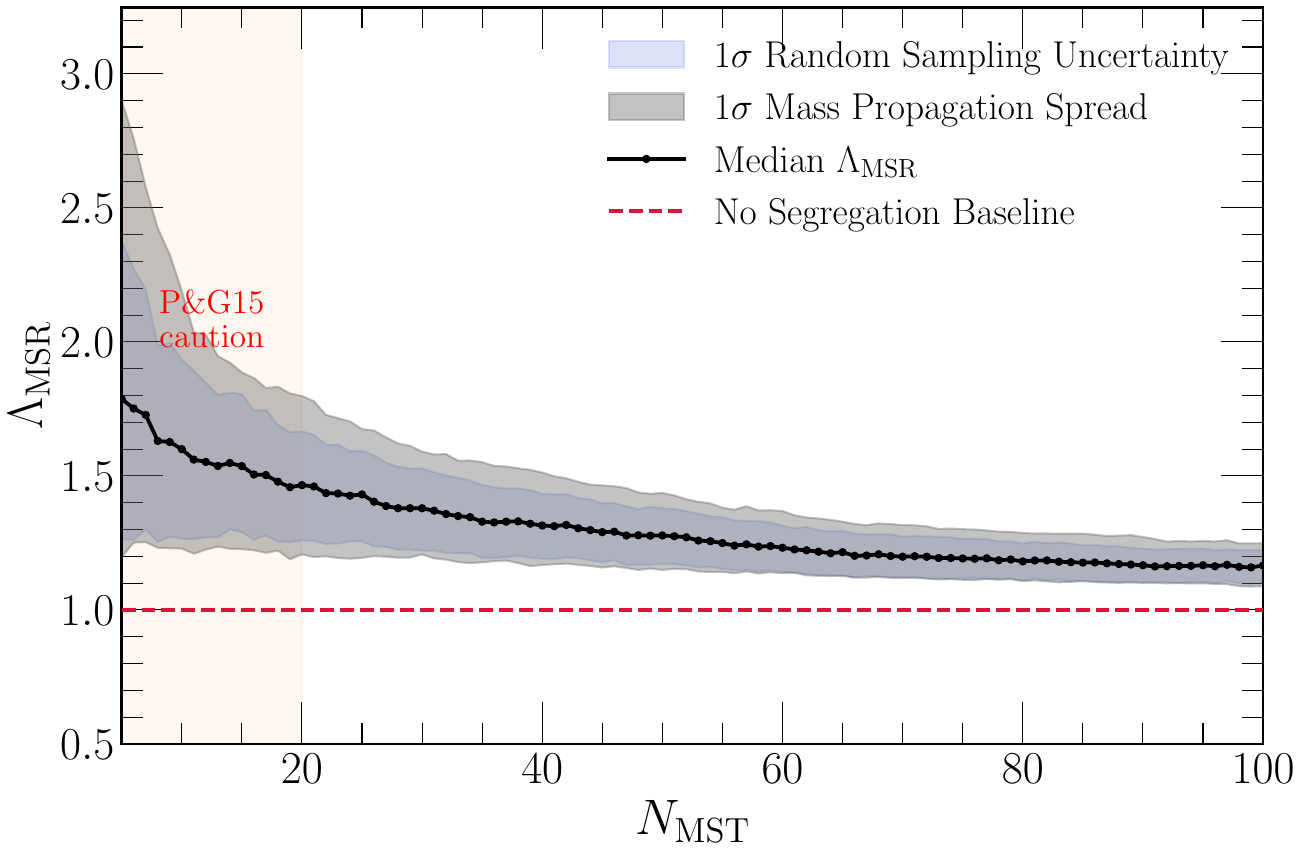}
	\caption{mass segregation ratio $\Lambda_{\text{MSR}}$ as a function of the number $N_{\text{MST}}$ of the most massive members. The geometric caution zone where $N_{\text{MST}} < 20$ is shaded, following \citet{Parker:2015}. The solid black line is the seed-averaged median $\Lambda_{\text{MSR}}$ over the Monte Carlo mass iterations, formed as the mean of the medians of four master seeds; the shaded bands show the $1\sigma$ random-sampling uncertainty (light blue) and mass-propagation spread (grey). The red dashed line marks the no-segregation baseline.}
	\label{fig:segregasi_massa}
\end{figure*}

The calculated median $\Lambda_{\text{MSR}}$ (solid black line in Figure \ref{fig:segregasi_massa}) remains above unity across the entire probed range ($N_{\text{MST}} = 5$ to $100$, the upper limit of this analysis). We therefore report the full curve but interpret only the portion with $N_{\text{MST}} \geq 20$, following the topological cautions of \citet{Parker:2015}; the values below that limit are shown for completeness and are excluded from all dynamical inference.

The leftmost region of the plot ($N_{\text{MST}} < 20$, highlighted in yellow) exhibits the highest absolute segregation ratios, whose seed-averaged median reaches $\Lambda_{\text{MSR}} = 1.79$ at $N_{\text{MST}} = 5$, the individual seeds spanning $1.62$--$1.94$. In this low-$N$ limit the MST topology is highly volatile: the chance clustering of a very small number of massive systems, for instance an unresolved dense pair near the cluster centre, can inflate $\Lambda_{\text{MSR}}$ in a way that carries no dynamical meaning. This regime is therefore treated strictly as a geometric caution zone and excluded from the interpretation above. 

Beyond the statistical caution zone, the dynamical signature of the cluster becomes apparent. From $N_{\text{MST}} = 20$ to $N_{\text{MST}} = 100$, the median $\Lambda_{\text{MSR}}$ declines overall from $\sim 1.47$ to $\sim 1.17$, with point-to-point fluctuations of order the quoted uncertainty. The steepest part of that decline, from the seed-averaged $1.79$ at $N_{\text{MST}} = 5$ to $\sim 1.47$ at $N_{\text{MST}} = 20$, occurs entirely within the small-number caution zone and is therefore not interpreted. Across the interpreted range the ratio remains above unity at every step, and both the random-sampling and mass-propagated lower bounds stay strictly above $\Lambda_{\text{MSR}} = 1$ throughout, so the segregation signal cannot be attributed to spatial sampling or to mass-estimation uncertainty. The persistence of the signal across such a large fraction of the cluster population indicates a pervasive mass segregation rather than localized sub-clustering, and its gradual weakening with increasing $N_{\text{MST}}$ is the expected behaviour of a stellar system evolving toward energy equipartition through two-body relaxation \citep{Spitzer:1987, Binney:2008}.

For an intermediate-age open cluster such as NGC 6134, the half-mass relaxation time ($t_{\text{rh}}$) has been measured from Gaia kinematics and found to be of order tens of millions of years, with \citet{Tarricq:2022} reporting a comparable value and \citet{Zeng:2025} obtaining $t_{\text{rh}} = 16.59 \pm 0.25$~Myr for this cluster. We adopt these published values rather than deriving $t_{\text{rh}}$ here, since doing so would require both a velocity dispersion, which this study does not measure, and masses for the full 890-star membership, which ASteCA v0.4.3 provides only for the 655 inner members. Since the cluster age ($\sim 1.4$~Gyr) exceeds $t_{\text{rh}}$ by nearly two orders of magnitude, the observed segregation is consistent with an origin in long-term dynamical evolution rather than in primordial formation conditions.

\begin{figure*}[t!]
	\centering
	\includegraphics[width=0.95\textwidth]{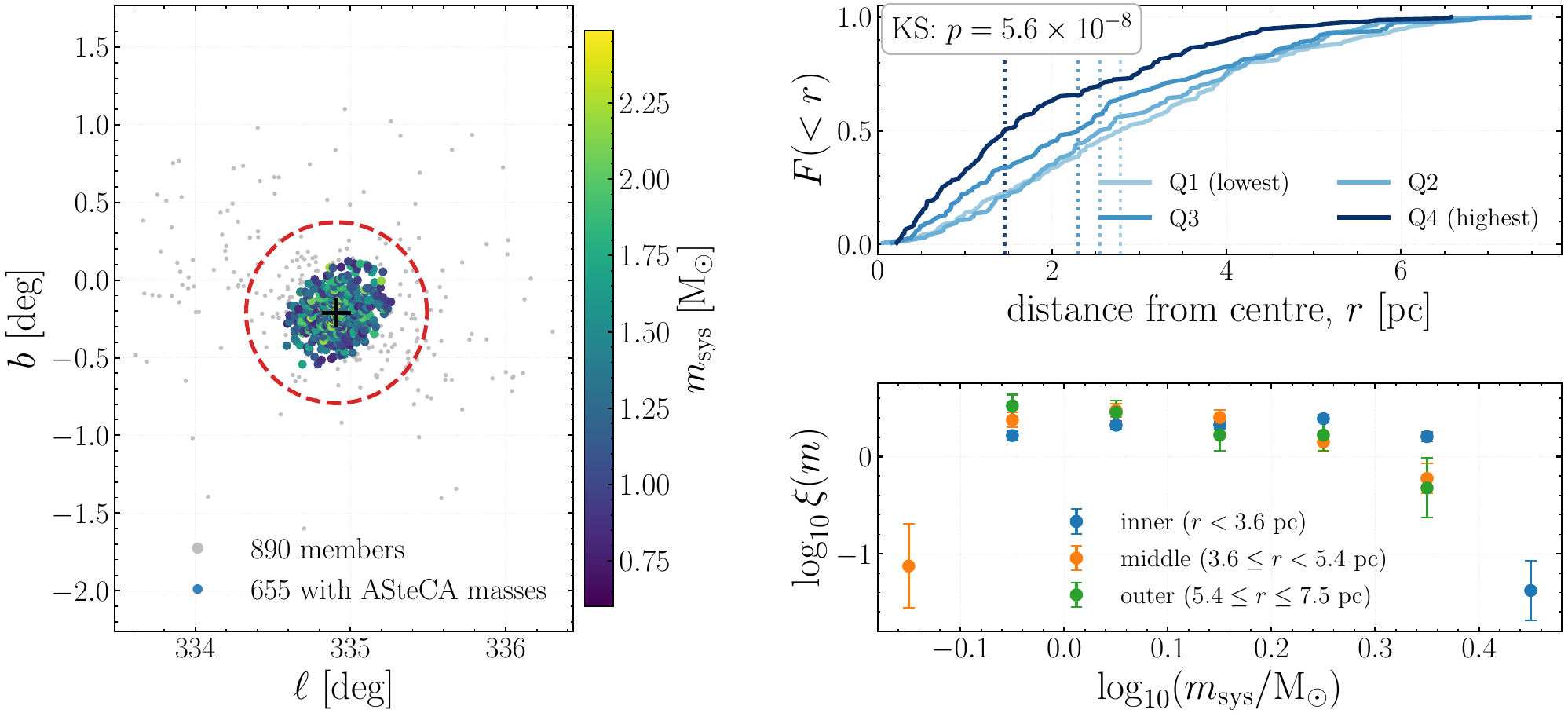}
	\caption{Complementary mass-segregation diagnostics of NGC~6134, computed on the 655 members for which ASteCA v0.4.3 assigns masses. Left: tangent-plane distribution of the members color-coded by the expected system mass $m_\mathrm{sys} = m_1 + p_\mathrm{bin}m_2$; the cross marks the fitted cluster center and the dashed circle the line-of-sight King core scale ($\mathrm{scl} = 10.9$~pc) shown for reference, a scale length along the line of sight that is not itself a projected radius; the $655$ members, plotted here at their projected radii, all lie within it, and ASteCA masses reach only $7.5$~pc. Upper right: cumulative radial distribution of the members in the four mass quartiles; the dotted vertical lines locate the median radius of each quartile, with the bootstrap $95\%$ confidence intervals given in the text. Lower right: present-day mass function (PDMF), $\xi \equiv \mathrm{d}N/\mathrm{d}\log_{10}m$ (normalized per member) in three annuli bounded by $\mathrm{scl}/3$, $\mathrm{scl}/2$, and the $7.5$~pc radial reach of the v0.4.3 masses; error bars are Poisson.}
	\label{fig:mass_segregation_radial}
\end{figure*}

In addition to the MST-based $\Lambda_{\text{MSR}}$ analysis above, we provide complementary, direct diagnostics of mass segregation, illustrated in Figure \ref{fig:mass_segregation_radial}: the spatial distribution of the members color-coded by mass, their radial distribution in different mass bins, and the mass function evaluated in annuli centered on the cluster center. For consistency with the mass-function analysis, we adopt the expected system mass $m_\mathrm{sys} = m_1 + p_\mathrm{bin}\,m_2$ (Section \ref{subsec:asteca-params}) and the projected cluster center fitted from the member $\ell$ and $b$ distributions, $(\ell_\mathrm{c}, b_\mathrm{c}) = (334.91^\circ, -0.21^\circ)$, and convert angular separations to physical radii using the per-member distance posterior modes of Section \ref{sec:distance}.

The left-hand panel of Figure \ref{fig:mass_segregation_radial} shows the tangent-plane distribution of the members, with the $655$ mass-bearing stars color-coded by $m_\mathrm{sys}$ and the remaining $235$ members of the $890$-star catalog plotted in gray for context. Two features of this panel set the scope of the mass-based analysis that follows. First, the mass-bearing sample is radially compact: their projected radii reach no further than $r = 7.5$~pc, which is below the line-of-sight King core scale $\mathrm{scl} = 10.9$~pc adopted in the distance prior (a scale length along the line of sight, shown for reference rather than as a projected radius, so the two are not directly comparable). Second, the mass-bearing sample is radially compact compared with the full catalog, whose projected radii extend to $r_{90} = 13.3$~pc and $r_\mathrm{max} = 27.9$~pc. The quantitative consequences of this restricted radial coverage are drawn out below.

The upper right-hand panel of Figure \ref{fig:mass_segregation_radial} shows the cumulative radial distribution of the members in the four mass quartiles. The median projected radius decreases from $2.79$~pc for the least massive quartile to $1.46$~pc for the most massive quartile, whose bootstrap $95\%$ confidence intervals ($[2.39, 3.28]$ pc and $[1.26, 1.77]$ pc, respectively) do not overlap; the two intermediate quartiles are statistically indistinguishable ($2.55$ and $2.29$~pc). The most massive $10\%$ of the members ($66$ stars, $m_\mathrm{sys} \geq 2.06\,M_\odot$) have a median radius of only $1.24$~pc, compared with $2.32$~pc for the full sample. Two-sample Kolmogorov--Smirnov (KS) tests confirm that the radial distribution of the most massive quartile differs highly significantly from that of the least massive one ($p = 5.6 \times 10^{-8}$), and that members with $m_\mathrm{sys} > 1.4\,M_\odot$ are significantly more centrally concentrated than those with $m_\mathrm{sys} \leq 1.0\,M_\odot$; the Mann--Whitney $U$ test yields an equivalent result for the extreme quartiles ($p = 4.4 \times 10^{-9}$). These conclusions are robust to the stricter membership threshold $P \geq 0.8$ (KS test between the extreme quartiles: $p = 8.5 \times 10^{-6}$).

Finally, the lower right-hand panel of Figure \ref{fig:mass_segregation_radial} presents the mass function in three projected annuli bounded by $r = \mathrm{scl}/3$, $r = \mathrm{scl}/2$, and the radial reach of the masses, where $\mathrm{scl} = 10.9$~pc is the scale length of the line-of-sight King distance prior (Section \ref{sec:distance}) and serves here only as a convenient reference scale for defining the annuli; the radii themselves are projected. ASteCA v0.4.3 assigns a mass to only the 655 members that survive its photometric decontamination, and those members reach no further than $r = 7.5$~pc; the three annulus boundaries are $r < 3.6$~pc, $3.6 \leq r < 5.4$~pc, and $5.4 \leq r \leq 7.5$~pc, containing 479, 134, and 42 members respectively. Over this range the fraction of systems with $m_\mathrm{sys} > 1.4\,M_\odot$ declines monotonically with radius, from $53\%$ in the innermost annulus to $33\%$ and then $31\%$ in the two outer ones, which is the expected signature of mass segregation acting over the region where masses exist. A direct comparison of the radial distributions of the inner and outer parts of this range confirms the same trend: the median projected radius rises from $2.12$~pc for the $613$ members within $r < 5.4$~pc to $5.91$~pc for the $42$ members in the outer annulus, and the Mann--Whitney $U$ test yields $p = 9.9 \times 10^{-28}$.

This measurement therefore describes the cluster core and inner envelope only. The projected mixture-model decomposition discussed in Section \ref{subsec:gmm}, which uses stellar positions alone and is consequently independent of the mass tables, separately resolves an elongated tail whose median projected radius is $11.5$~pc with a $90^\mathrm{th}$ percentile radius of $20.4$~pc, both well beyond the $7.5$~pc reach of the masses. The two analyses thus sample disjoint regions of the cluster, and the mass-based analysis above cannot be used to trace segregation into the tidal structure. We therefore make no claim regarding the mass composition of the tail; establishing that would require mass estimates extending beyond the radial reach of the v0.4.3 decontaminated sample.

\subsection{The Mass Function of NGC 6134}\label{subsec:pdmf}
We express the present-day mass function (PDMF) in logarithmic form, $\mathrm{d}N/\mathrm{d}\log_{10} m \propto m^{-\Gamma}$, so that the slopes relate to the customary power-law index by $\Gamma = \alpha - 1$ (this $\alpha$ is the mass-function power-law index, distinct from the binary-probability coefficient $\alpha$ of Section~\ref{subsec:asteca-params}). In this convention the canonical Salpeter value is $\Gamma = 1.35$ \citep{Salpeter:1955}, and the widely adopted universal IMF is a three-segment power law with $\Gamma \approx 1.3$ above $\sim 0.5\,M_\odot$, $\Gamma \approx 0.3$ at intermediate mass, and $\Gamma \approx -0.7$ at low mass \citep{Kroupa:2001}. Most clusters follow this form closely \citep{Cordoni:2023}, but departures are expected: two-body relaxation preferentially strips the least massive members, flattening the low-mass end, while stellar evolution removes the most massive stars. The sense and amplitude of this reshaping are quantified by direct N-body simulations, which find the evolution of the slopes to be strongest for intermediate-mass stars and at later evolutionary stages \citep{Baumgardt:2023}; we compare our results with those predictions in Section \ref{sec:compare}. The comparison of our low-mass segment with these canonical slopes is deferred to Section \ref{sec:compare}, because the lowest-mass bins of our fit are populated by unresolved companions rather than by directly counted primaries.

We therefore derive the PDMF of NGC 6134, the distribution inferred from present-day observations, and ask where it departs from the canonical form.

We construct the PDMF from the ASteCA mass estimates of Section \ref{subsec:asteca-params}, in which each member's mass follows from its $G$ magnitude and $(G_\mathrm{BP} - G_\mathrm{RP})$ color via the best-fit isochrone. Because NGC 6134 has a binary fraction $f_{\mathrm{bin}} \approx 0.4$, we include both components for members flagged as binaries, giving each system the expected mass $m_{\mathrm{sys}} = m_1 + p_{\mathrm{bin}} m_2$. The $20$ independent ASteCA realisations are treated as an empirical sampling distribution of the cluster's PDMF. We emphasise one limitation that governs the interpretation of the low-mass bins: ASteCA assigns no primary with $\log_{10}(M/M_\odot) \lesssim -0.2$, so below that limit the distribution is carried by binary companions weighted by $p_{\mathrm{bin}}$ together with the low-mass tail of the mass-error draws. The steep rise below $-0.2$ therefore traces the growth of the companion contribution, not a physical feature of the single-star mass function.

We apply no analytical completeness correction, and instead restrict the fit to the mass range over which the census of members is essentially complete. All primaries in the 655-member sample fall within $10.6 \lesssim G \lesssim 18.0$~mag, where the Gaia DR3 catalogue is close to complete ($\sim 100\%$ between $G \simeq 12$ and $17$, declining only at fainter magnitudes \citep{Riello:2021, GaiaDR3:2023}). In mass space these primaries span $\log_{10}(M/M_\odot) \gtrsim -0.15$ ($M \gtrsim 0.7\,M_\odot$); the faint limit of the sample is set by the membership selection of \citet{Yusuf:2024} rather than by Gaia, so no primary-mass dependence of the completeness is expected inside the fitted range. The broken power law is therefore fitted over $\log_{10}(M/M_\odot) \in [-0.60, 0.30]$.

We do not weight the bins by membership probability. The sample already passes a two-stage selection: the HDBSCAN kinematic cutoff ($P \geq 0.5$) within the 30~pc aperture of Section \ref{subsec:membership}, followed by ASteCA's treatment of the cluster region, which restricts to its fitted cluster radius and then applies cell-based photometric cleaning of the color--magnitude diagram, discarding the lowest-probability star in each photometric cell in proportion to the local field density (Section \ref{subsec:asteca-params}). This leaves 655 members whose summed HDBSCAN membership probability is $N_{\mathrm{eff}} = 627$, only $4.2\%$ below the raw count. As an explicit test, reweighting the binned PDMF by the per-star membership probability changes the inferred parameters by $\Delta\Gamma_A = +0.04$, $\Delta\Gamma_B = +0.02$, and $\Delta\log_{10} M_c = +0.008$~dex, all well inside the quoted uncertainties, so we adopt the unweighted hard-count PDMF as fiducial.

\begin{figure}[t!]
	\centering
	\includegraphics[width=0.48\textwidth]{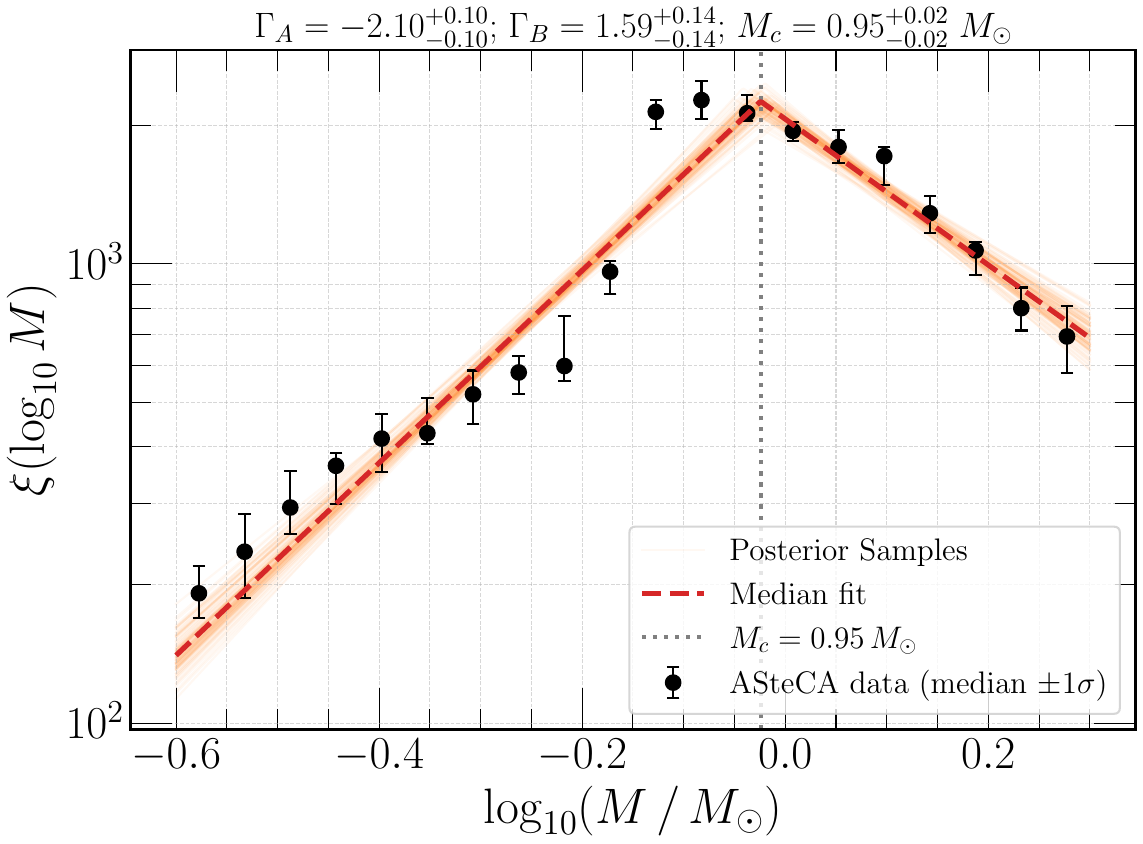}
	\caption{Present-day mass function (PDMF) of NGC 6134 derived from the ASteCA v0.4.3 mass estimates of the 655 decontaminated members. The broken power law gives a break at $M_c = 0.95^{+0.02}_{-0.02}\,M_\odot$ ($\log_{10}(M/M_\odot) = -0.024^{+0.009}_{-0.008}$), separating a low-mass regime with slope $\Gamma_A = -2.10^{+0.10}_{-0.10}$ from a high-mass regime with slope $\Gamma_B = 1.59^{+0.14}_{-0.14}$. The fit is obtained by nested sampling on $20$ bins of width $0.045$~dex over $\log_{10}(M/M_\odot) \in [-0.60, 0.30]$. The nine lowest-mass bins, below $M = 0.64\,M_\odot$, contain no directly counted primaries in any of the $20$ realisations and are populated entirely by unresolved companions (see text).}
	\label{fig:pdmf-fit}
\end{figure}

The broken power law is fitted by nested sampling with the Nautilus package \citep{Lange:2023}, using $20$ bins of width $0.045$~dex over $\log_{10}(M/M_\odot) \in [-0.60, 0.30]$ and $1000$ live points. All random draws, including the per-star Gaussian mass-error draws, use fixed seeds, so the quoted values are exactly reproducible; we additionally verified that the fitted parameters are stable against the choice of seed. The resulting best-fit model is shown in Figure \ref{fig:pdmf-fit}.

The PDMF of NGC 6134 departs measurably from a canonical IMF: the low-mass slope is negative ($\Gamma_A = -2.10^{+0.10}_{-0.10}$, i.e., the distribution rises toward higher mass throughout the fitted range), the high-mass slope is steeper than the canonical Salpeter value ($\Gamma_B = 1.59^{+0.14}_{-0.14}$), and the break mass ($M_c = 0.95^{+0.02}_{-0.02}\,M_\odot$, or $\log_{10}(M_c/M_\odot) = -0.024^{+0.009}_{-0.008}$) lies above the characteristic turnover mass of the standard IMF. Because the low-mass segment is carried by unresolved companions, the offset in $M_c$ indicates a relative scarcity of sub-solar-mass systems rather than a reconstruction of the single-star IMF; a detailed comparison with the universal IMF and with N-body predictions is given in Section \ref{sec:compare}.

It is also worth stating what mass this analysis actually samples. Summing the primary masses of the members whose primary falls within the fitted mass window gives $724 \pm 15\,M_\odot$, and summing their expected system masses gives $902 \pm 21\,M_\odot$, where the uncertainties are the standard deviations across the $20$ realisations. These values describe the decontaminated sample only, not the cluster as a whole: the members whose primary mass lies above the upper limit of the fit window, the $235$ catalogued members that lie beyond the radius at which v0.4.3 assigns masses, and all stars fainter than the spectroscopic limit of the sample, are absent from the sums. We therefore do not quote a total cluster mass. Integrating the fitted power law downward does not recover the missing population, because its low-mass segment is dominated by binary companions that are already included in the $655$ counted here; that integral both overpredicts the star count and underpredicts the summed system mass, and so is not a measure of the cluster total.

It is worth stating the mass range over which the fit is constrained. The $655$ primaries span $0.69 \leq M_1/M_\odot \leq 2.15$, so the nine lowest-mass bins of the fit window, below $M = 0.64\,M_\odot$, contain no directly counted primary in any of the realisations and are populated entirely by unresolved companions; $\Gamma_A$ is therefore the mass distribution of system components below $M_c$, not a reconstruction of the primary mass function, and no member is detected as a primary below $0.69\,M_\odot$. At the upper end, the fit window terminates at $M = 2.00\,M_\odot$, so that between $0$ and $32$ of the $655$ most massive members (median $13$) lie outside the fit. We verified that this truncation does not bias $\Gamma_B$: excluding the highest-mass bin shifts it by $0.04$ and extending the window to admit all members shifts it by $0.03$, both well within the fitted uncertainty of $0.14$.

\subsection{Substructure detection using Gaussian Mixture Models (GMM)}\label{subsec:gmm}
Several of our inferences rest on assuming that the cluster is spherically symmetric: the King profile used for the distance prior, the projected density profile of Section \ref{subsec:structure}, and the assumption of a common distance implicit in the radial mass analysis. We therefore test that assumption directly, by decomposing the stellar positions of the cluster into substructures without reference to any mass or velocity information. Substructures in OCs are common, as demonstrated by the work of \citet{Trullols:1997}, who identified new substructures in IC 348 using UBVRI photometric data, and by \citet{Tarricq:2022}, who successfully detected the presence of tidal tails in 71 open clusters using the GMM method on 2D projection data from Gaia EDR3. The GMM is a clustering algorithm that assumes the data distribution can be represented as a mixture of one or more Gaussian components. It fits the observed data using the Expectation Maximization (EM) technique, which computes the probability that each cluster member belongs to each of the Gaussian components, the number of which is set by the user. The GMM algorithm is well-suited for detecting substructures in a stellar cluster because it can distinguish different density distributions even when the component distributions overlap. Another advantage of the GMM is that it does not require the cluster members to follow a circular distribution in the plane of the sky, an assumption strictly used by the HDBSCAN algorithm. 

In this work, stellar positions were transformed from equatorial coordinates ($\alpha,\delta$) into local tangent-plane coordinates using a gnomonic projection centered on the cluster. Because the angular extent of the system is small, distortions introduced by the projection are negligible compared with the characteristic scales of the substructures investigated. The GMM was therefore applied in the projected Cartesian coordinate system (x,y).

To map the spatial architecture of NGC 6134 beyond a spherically symmetric approximation, we model the 2D and 3D projected positions of the cluster members using a Dirichlet Process Gaussian Mixture Model (DPGMM). Unlike standard GMMs requiring a fixed number of clusters, the Dirichlet Process prior determines the number of components from the data density alone, without a fixed $k$.

While Dirichlet Process models are theoretically non-parametric, finite truncations are prone to artificially over-fragmenting single coherent physical structures to accommodate natural non-Gaussianities in spatial distributions (e.g., \citealp{Miller:2013}). To capture complex astronomical morphology without overly restricting the model's flexibility, we set a conservative truncation limit of $N_{\text{components}} \leq 8$ and a Dirichlet concentration $\alpha_\mathrm{DP} = 0.01$, which controls the expected sparsity of the component weights. Varying $\alpha_\mathrm{DP}$ from $0.001$ to $1.0$ leaves the distribution of active-component counts unchanged across $30$ random seeds, so the decomposition below does not depend on this choice. Rather than relying strictly on the raw mixture components as distinct physical entities, we map these latent variables to physically interpretable structural features: Core, Halo, and Tails, by applying a rigorous, domain-informed classification hierarchy:
\begin{itemize}
	\item Core: Defined as the component displaying the highest peak surface-density proxy, representing the densest, most compact, and gravitationally dominant region of the cluster.
	\item Halo: Physically modeled as a diffuse, concentric envelope. A component is classified as a Halo candidate if it is more extended than the Core ($\text{R}_{\text{eff, halo}} > \text{R}_{\text{eff, core}}$) and shares a significant spatial overlap with it, such that their centroids are separated by less than $0.5 \times (\text{R}_{\text{core}} + \text{R}_{\text{self}})$. If multiple components satisfy these criteria, the most extended one is selected. In the absence of a qualified candidate (e.g., a highly disrupted cluster possessing only tidal features without a diffuse envelope), the algorithm falls back to assigning the spatially closest non-core component as the Halo, flagging a user warning for manual inspection.
	\item Tails: All remaining active components are interpreted as structural perturbations or tidal features if they are genuinely elongated, i.e., with an axis ratio $b/a < 0.5$, or spatially displaced from the Core by more than $0.5 \times (\text{R}_{\text{core}} + \text{R}_{\text{self}})$.
	\item Any component left over (round, concentric with the Core, neither Tail nor Halo) is a spurious scale-split of the Core envelope and absorbed into the Core classification. This ensures that the final set of components is physically interpretable and avoids over-fragmentation of the Core.
	
\end{itemize}

We emphasize that this three-component decomposition is not imposed a priori. In an earlier version of the analysis, the initial 2D projection occasionally returned four active components; close inspection revealed that the fourth component was a purely concentric scale-split of the inner envelope, a known over-fragmentation effect of finite Dirichlet Process truncations \citep{Miller:2013}. We therefore merge purely concentric components through an exact mixture-moment transform before classification, which yields a robust, physically distinct three-component model: Core, Halo, and Tail, in both the 2D and the 3D fits, resolving the apparent discrepancy between earlier figures. A two-component solution is not excluded by the data: fixed-$k$ selection returns $k=2$ as the best single description of the projected positions, although $k=4$ and $k=5$ lie within $\Delta\mathrm{BIC}=2$ of it and the Dirichlet Process favours four or more active components in $87\%$ of seeds. Component counting is in any case blind to whether a component is physically distinct, which is the criterion applied here: fitting the same pipeline with two components does not remove the elongation but merges the tail into a broader halo, diluting its axis ratio from $b/a=0.32$ to $b/a=0.68$, whereas the three-component model isolates a tail whose elongation is significant against isotropic mocks at $p=0/30$. The validity of this elongation is not asserted; it is demonstrated by the size-matched isotropic null test (Section~\ref{subsec:isotropy-null}), which shows that isotropic realizations of an equally sized sample never produce a component as elongated as the observed Tail.

\begin{figure}[!t]
	\centering
	\includegraphics[width=0.48\textwidth]{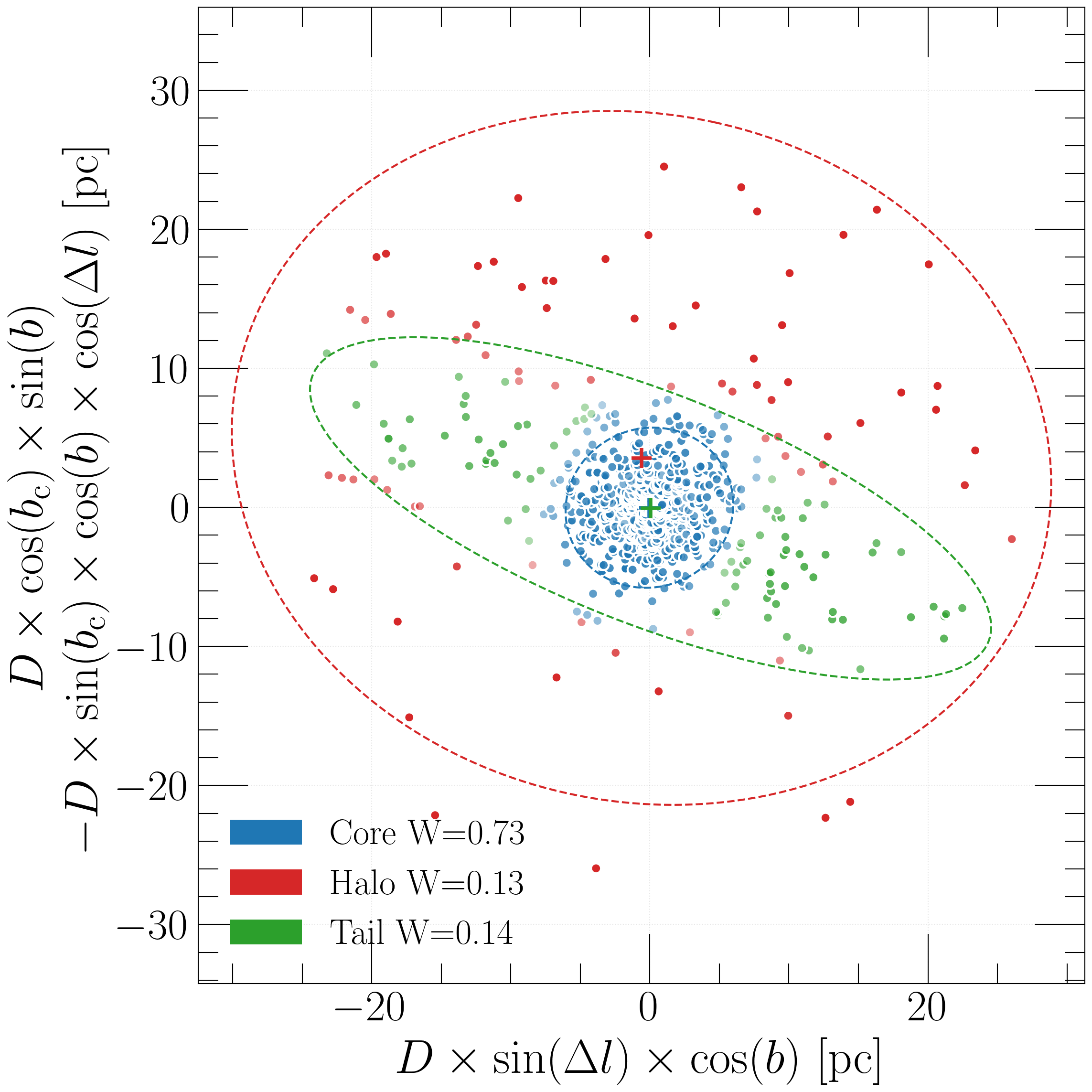}
	\caption{Structural decomposition of NGC 6134 using a multi-component Gaussian mixture. The distribution is partitioned into a central Core (blue), a diffuse Halo (red), and a Tail (green), classified according to the hierarchical density and scale criteria described in Subsection~\ref{subsec:gmm}. Ellipses represent the $95\%$ Credible Intervals (equivalent to $2\sigma$) covariance for each identified component. Component weights are provided in the legend.}
	\label{fig:hasil_gmm}
\end{figure}

The GMM algorithm has a stochastic nature, meaning that every iteration returns a slightly different result. We therefore iterate the GMM a thousand times to achieve a convergent result. The final Gaussian parameters of each component are taken as the mode of all the individual iteration results, with the standard error computed as $\mathrm{MAD}/\sqrt{N_\mathrm{iter}}$, where $N_\mathrm{iter} = 1000$ is the number of iterations. Our GMM final results (Figure \ref{fig:hasil_gmm}) show that NGC 6134 can be fitted with three Gaussian components. The final assignment of cluster members to a component is based on the component that most frequently includes them as members. The members of each component are shown with a different color in Figure \ref{fig:hasil_gmm}.

To characterize the identified components, we extract their geometric properties directly from the covariance matrix $\Sigma_k$ of each Gaussian component. An eigendecomposition of $\Sigma_k$ yields eigenvalues $\lambda_1 \ge \lambda_2$ and the associated eigenvectors; from these we define the effective RMS radius $r_\mathrm{rms} = \sqrt{\lambda_1 + \lambda_2}$ (i.e., the square root of the trace), the axis ratio $b/a = \sqrt{\lambda_2/\lambda_1}$ (a measure of elongation, with $b/a = 1$ for a round component), and the position angle $\mathrm{PA} = \operatorname{atan2}(v_{1,y}, v_{1,x})$ of the major axis, measured counterclockwise from the projected $x$-axis. As a proxy for peak surface density we adopt $w_k / \sqrt{\det\Sigma_k}$, which is used to identify the Core. The values for the three surviving components are listed in Table~\ref{tab:component-props}.

\begin{table}[!t]
	\centering
	\begin{threeparttable}
		\caption{Geometric properties of the three DPGMM components of NGC 6134 in the projected plane of sky: statistical weight $w$, number of assigned members $N$, effective RMS radius $r_\mathrm{rms}$, axis ratio $b/a$, and position angle $\mathrm{PA}$ of the major axis.}
		\label{tab:component-props}
		\begin{tabular}{lccccc}
			\toprule
			Component & $w$ & $N$ & $r_\mathrm{rms}$ [\si{pc}] & $b/a$ & $\mathrm{PA}$ [\si{deg}] \\
			\midrule
			Core & $0.73$ & 722 & $3.41$ & $0.94$ & $21.8$ \\
			Halo & $0.13$ & 78 & $15.77$ & $0.83$ & $167.6$ \\
			Tail & $0.14$ & 90 & $11.20$ & $0.32$ & $158.5$ \\
			\bottomrule
		\end{tabular}
	\end{threeparttable}
\end{table}

This decomposition is based on stellar positions alone and is therefore independent of the mass estimates used in Section~\ref{subsec:mass-segregation}. The Core, Halo, and Tail reported here correspond to median projected radii of $2.5$, $18.1$, and $11.5$~pc respectively, converted from the tangent-plane coordinates at the adopted cluster distance of $1070.8$~pc (Section~\ref{sec:distance}); this is a common reference distance, whereas the radial mass analysis of Section~\ref{subsec:mass-segregation} uses the individual Bayesian distances of the members. The Core overlaps the region in which ASteCA v0.4.3 assigns masses while the Halo and Tail lie almost entirely beyond the $7.5$~pc reach of those masses. We consequently restrict the mass-based discussion to the Core and make no claim about the mass composition of the other two components.

The asymmetric Tail is not accessible to the King profile used for the distance estimation (Section~\ref{sec:distance}): that model is one-dimensional and spherically symmetric, providing a robust gravitational centre of mass and a physical tidal-cutoff radius, but it cannot represent asymmetric spatial elongations. The mixture model is therefore required precisely to map these asymmetries that the King profile inherently misses.

We also applied the mixture model to the three-dimensional distribution of the members. As established in Section \ref{subsec:structure}, the distance distribution is not sharply defined, individual members deviating from the cluster distance by enough to produce an elongated rather than concentric morphology in the projected panels of Figure \ref{fig:3d-spatial-distribution}. Because distance is inversely proportional to parallax, symmetric parallax errors translate into highly asymmetric distance distributions, so this line-of-sight stretching is an artifact rather than a physical elongation. The artifact is compounded by our kinematic membership determination, in which reliance on the same parallaxes alongside the proper motions admits stars with poorly constrained distances. Two mitigations are possible: stricter filtering of the Gaia query, for example limiting $f$, reduces the line-of-sight scatter at the cost of excluding fainter members; and introducing spatial constraints could isolate the dense core, provided one accounts for the risk of artificially truncating genuine plane-of-sky tidal features. The GMM is run a thousand times and the mode taken, to guard against the stochastic nature of the algorithm. We stress that the elongated component recovered in this three-dimensional run is a line-of-sight distance-dispersion structure, not the plane-of-sky tidal feature identified in the projected decomposition of Section \ref{subsec:gmm}; the two are distinct, and only the latter is tested against the null model below.

\begin{figure}[!t]
	\centering
	\includegraphics[width=0.48\textwidth]{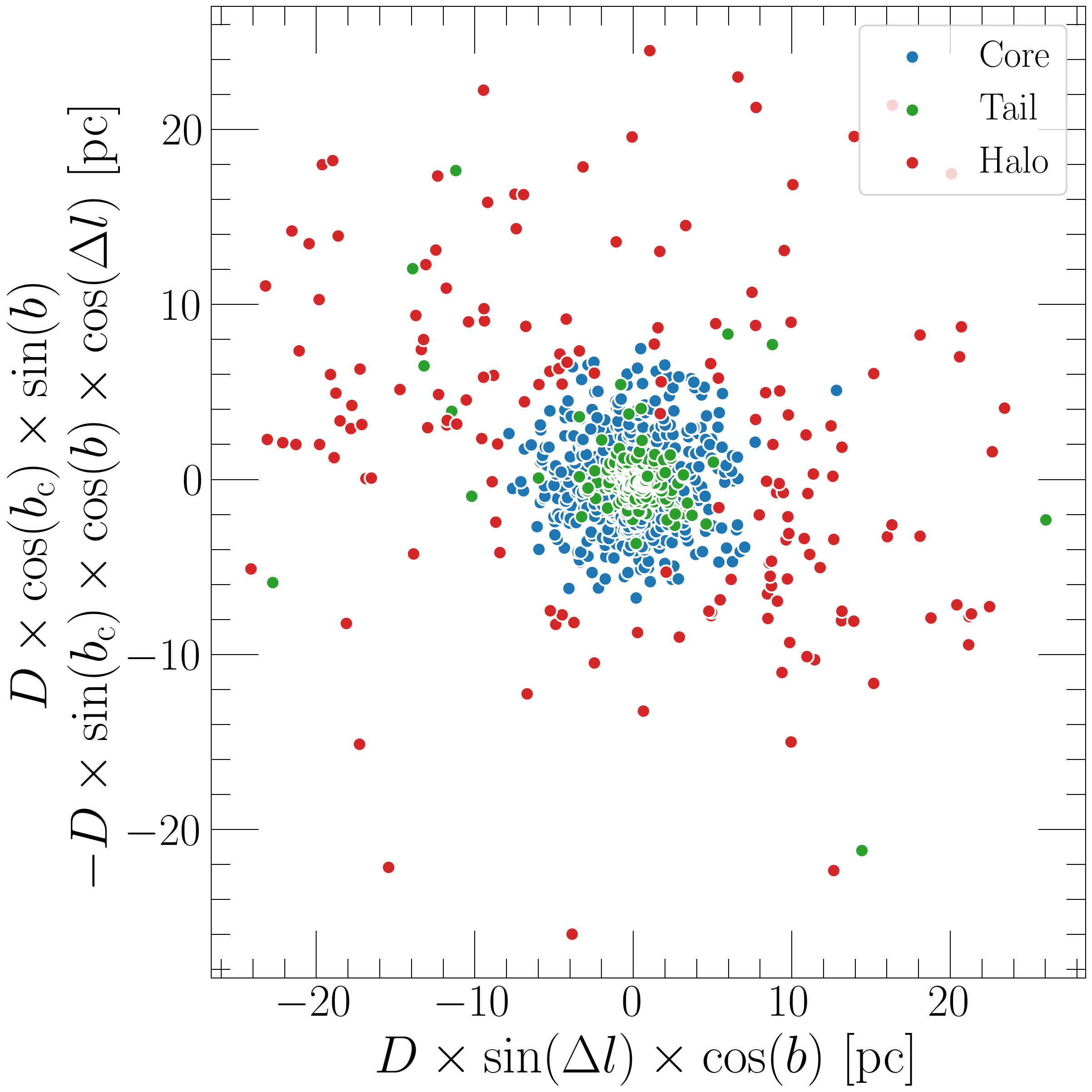}
	\caption{The GMM clustering results on the 3D distribution of the cluster members of NGC 6134, shown in the projected tangent plane, with each substructure represented by a distinct color.}
	\label{fig:GMM_to_3D}
\end{figure}

\begin{figure*}[t]
	\centering
	\includegraphics[width=0.7\textwidth]{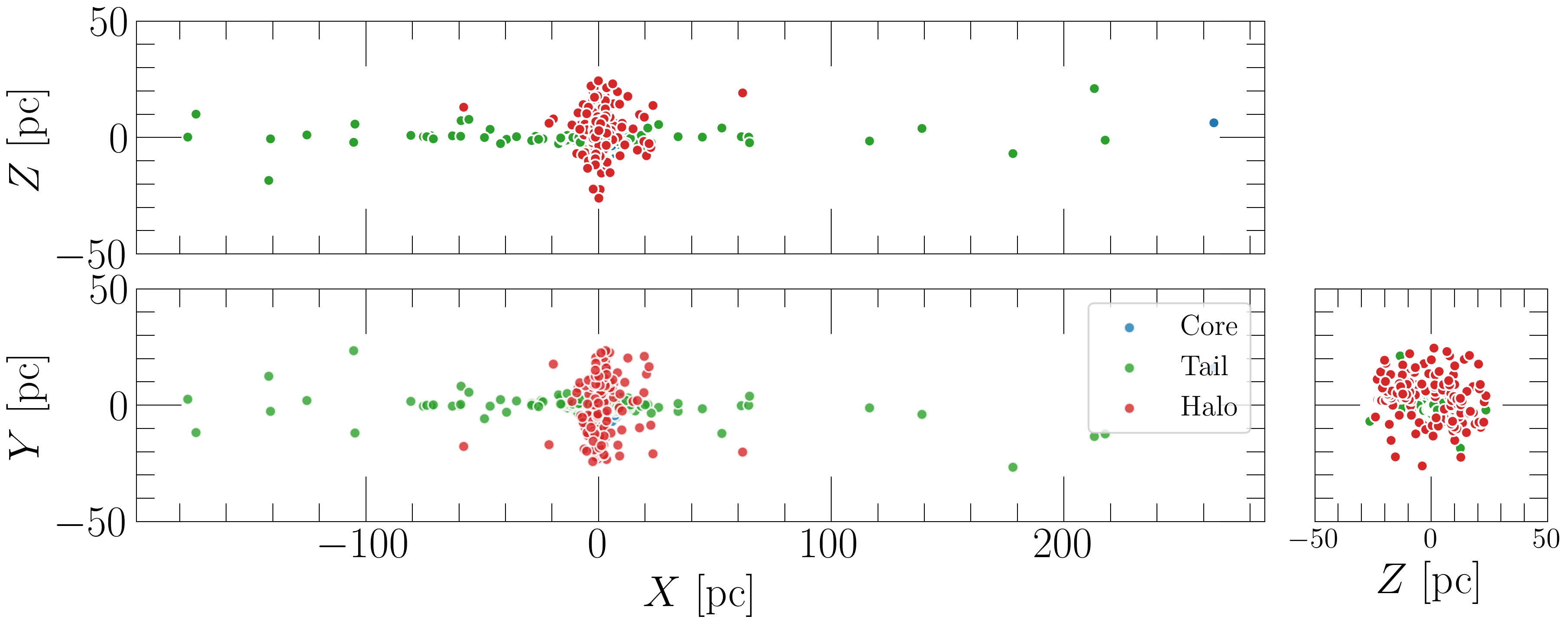}
	\caption{The GMM clustering results on the 3D data are projected onto the XY, YZ, and XZ planes, with each substructure represented by a distinct color.}
	\label{fig:substructures-xy-yz-xz}
\end{figure*}

\Needspace{4\baselineskip}%
\subsection{Isotropy null test}\label{subsec:isotropy-null}

To quantify whether the elongated tail identified in the projected decomposition (Figure \ref{fig:hasil_gmm}) could arise purely from sample noise, we performed a size-matched isotropy null test. The test targets this two-dimensional tail specifically; the elongated component seen in the three-dimensional run is a separate line-of-sight structure and is not subject to this comparison. We generated 30 independent, isotropic realizations of the NGC 6134 member positions with the same number of stars ($N = 890$) and matched the mock clouds' spread to the observed halo ($r_{\mathrm{rms}} = 15.77$~pc, i.e., a per-coordinate scale of $15.77/\sqrt{2} \approx 11.15$~pc), and processed them through the same DPGMM settings used for the data ($\alpha_\mathrm{DP} = 0.01$, maximum of eight components, full covariance, random initialization with $n_{\mathrm{init}} = 50$, identical survival weight filter $w \ge 0.1$, same convergence tolerance). For each realization we retained the most-elongated surviving component, i.e., the smallest axis ratio $b/a$ among the active components, thereby constructing the null distribution of this extreme statistic. If the tail were merely a spurious artifact of the Dirichlet process prior, which is known to occasionally over-fragment single physical structures \citep{Miller:2013}, these isotropic realizations should frequently yield a component as elongated as the observed one. 

\begin{figure*}[t]
	\centering
	\includegraphics[width=0.72\textwidth]{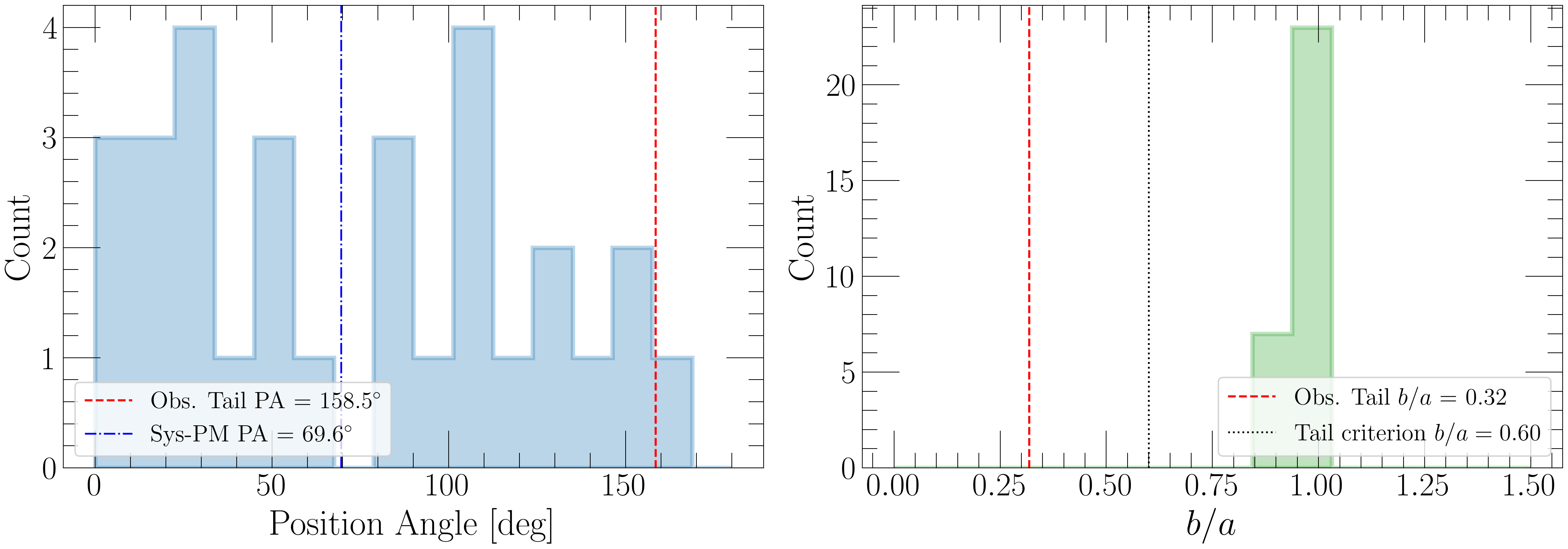}
	\caption{Isotropy null test validating the projected elongated tail of NGC 6134. Histograms of the most-elongated surviving component extracted by our DPGMM pipeline from 30 size-matched isotropic mock datasets ($N = 890$). Left: mock position angles (PA) are uniformly distributed, as expected for isotropic input; the observed tail ($158.5^\circ$, red dashed line) is nearly perpendicular to the cluster's systemic proper-motion direction ($69.6^\circ$, blue dash-dotted line; an $88.9^\circ$ offset). Right: mock axis ratios strongly favor near-circular components ($\langle b/a\rangle \approx 0.96$); the observed tail's elongation ($b/a = 0.32$, red dashed line) lies below every realization (0/30). An isotropic field therefore cannot reproduce the observed tail.}
	\label{fig:isotropy-null}
\end{figure*}

However, our results demonstrate otherwise. The observed Tail has an axis ratio of $b/a = 0.32$ at a position angle of $158.5^\circ$, whereas the most-elongated components produced by the 30 isotropic realizations scatter around $\langle b/a \rangle \approx 0.96$; none (0/30) reached an elongation as extreme as the observed tail (Figure~\ref{fig:isotropy-null}). Furthermore, the position angles of the mock components are uniformly distributed, as required for isotropic input, whereas the real tail lies nearly perpendicular (an $88.9^\circ$ offset) to the cluster's systemic proper motion; a specific, physically motivated orientation that the isotropic null never singles out. The extreme-value comparison therefore yields $\hat{p} = 0/30$ for reproducing the observed elongation by chance, with a one-sided 95\% upper bound of $\approx 9.5\%$. Combined with the coherent, systemically locked orientation of the tail, this strongly supports its physical reality rather than an algorithmic fitting artifact. The tail members additionally passed the same strict kinematic HDBSCAN membership cut ($P \ge 0.5$) applied in the three-dimensional proper-motion and parallax space used for membership (Section~\ref{subsec:membership}), so the feature cannot be attributed to residual field contamination.

\section{Comparison with Previous Results}\label{sec:compare}

\begin{figure*}[t!]
	\centering
	\includegraphics[width=0.7\textwidth]{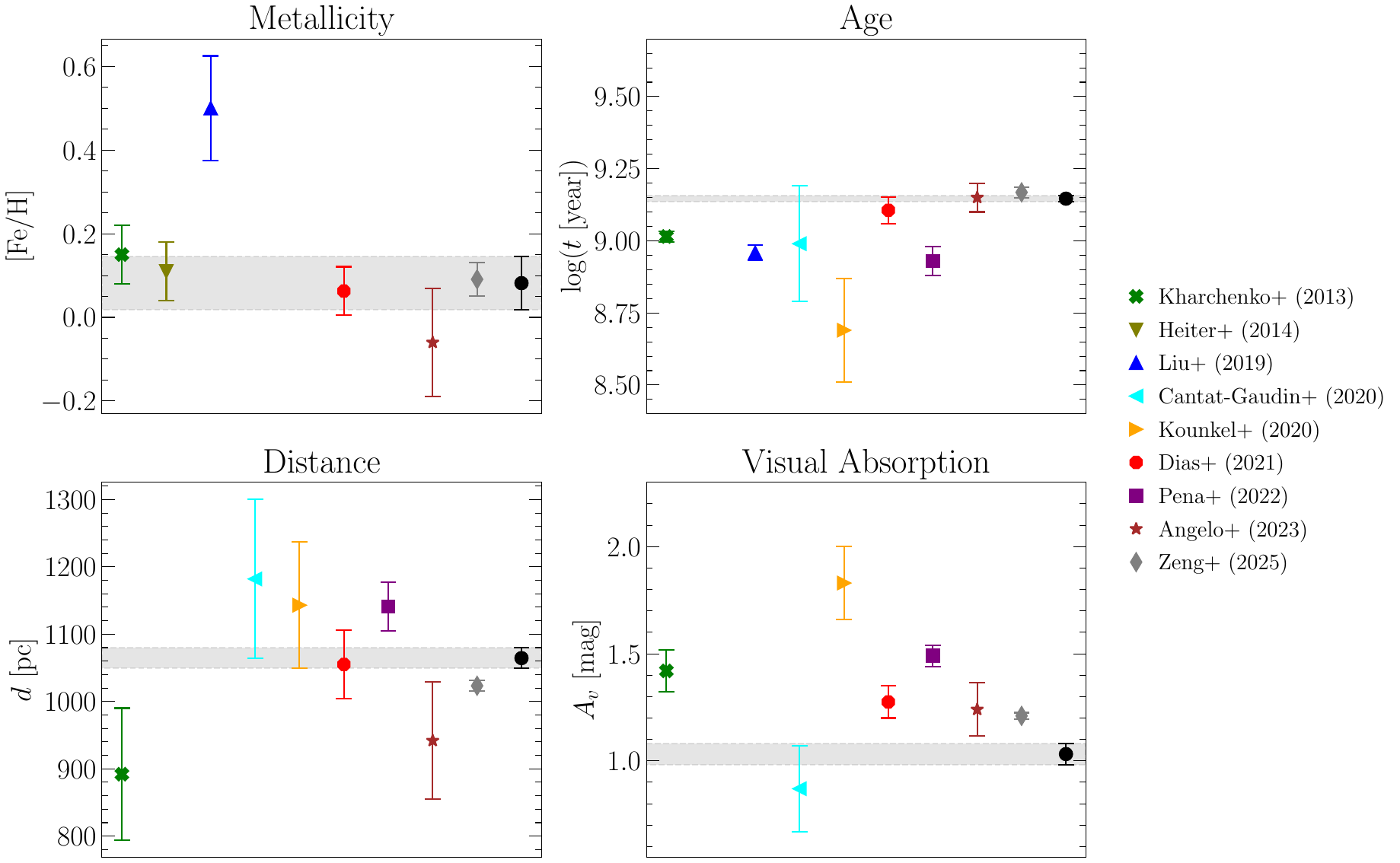}
	\caption{Comparison of the NGC 6134 cluster parameters $\mathrm{[Fe/H]}$, $\log(t)$, $d$, and $A_\mathrm{V}$ from this study (represented by black dots) with other studies. The shaded area indicates the parameter values obtained in this study.} 
	\label{fig:stellar-parameter-comp}
\end{figure*}

We compare the fundamental parameters derived in this study with those reported in ten previous studies spanning from 2013 to 2025, summarized in Table~\ref{tab:params_prev_study} and visualized in Figure~\ref{fig:stellar-parameter-comp}. The comparison sample includes one high-resolution spectroscopic study \citep{Heiter:2014}, obtained from the PASTEL catalog \citep{Soubiran:2010}, and nine photometric studies. Among the photometric works, five employed Gaia data at either DR2 or DR3 level \citep{LiuandPang:2019, CantatandAnders:2020, Kounkel:2020, Dias:2021, Angelo:2023}, two used infrared photometry from 2MASS \citep{Skrutskie:2006} and VISTA Variables in the Via Lactea \citep[VVV;][]{Minniti:2010} \citep{Kharchenko:2013, Pena:2022}, and two are recent independent Gaia DR3 analyses: \citet{Cavallo:2024} and \citet{Zeng:2025}. We restrict the comparison to studies that report formal uncertainties or
standard deviations.

\subsection{Membership sample}\label{subsec:compare-membership}

The membership determinations underlying these parameter estimates differ considerably, which partly explains the spread in reported values. \citet{Kounkel:2020} applied a clustering algorithm in five astrometric dimensions ($\ell, b, \mu_{\alpha^*}, \mu_\delta, \varpi$), while \citet{Dias:2021} and \citet{Angelo:2023} drew their membership lists from external catalogs. \citet{Zeng:2025} applied the same HDBSCAN algorithm used in this work, but confined their initial candidate selection to a $1\degr$ search radius, yielding 714 members with $P > 0.5$. Our previous membership study \citep{Yusuf:2024}, on which the present analysis is built, queried Gaia DR3 out to 90~pc from the cluster center ($4.35\degr$ at the reference distance of $1182$~pc \citep{Cantat-Gaudin:2020}), an aperture large enough that the cluster boundary was determined by the data rather than assumed: the cumulative count is complete within 30~pc, and the catalog we adopt is truncated at that radius, matching the tidal radius to which the King fit of \citet{Yusuf:2024} converges. We adopt the resulting 30~pc ($1.45\degr$) extent, $6.64$~deg$^2$, which is roughly twice the search area of \citet{Zeng:2025}. More significantly, their $1\degr$ aperture corresponds to $20.6$~pc and therefore truncates the cluster inside its measured boundary, excluding the $49$ members we find between $20.6$ and $29.8$~pc. Those peripheral members are systematically fainter and of lower membership probability than the inner population (median $G = 16.4$ versus $15.7$~mag; median $P = 0.92$ versus $0.999$), and their exclusion limits sensitivity to the low-density outer structure discussed in Section~\ref{sec:substructures}.

To quantify how different the two member lists are, we reproduced the \citet{Zeng:2025} selection on the same Gaia DR3 data (a $1\degr$ candidate cone with $\mathrm{RUWE}<1.4$ and $G\le19$, followed by HDBSCAN with \texttt{min\_cluster\_size}=6 and \texttt{min\_samples}=6 on $\mu_{\alpha^*}$, $\mu_\delta$, and $\varpi$; $P>0.5$). The reproduction returns 793 members, slightly more than the 714 reported by \citet{Zeng:2025}. Their exact candidate pre-selection is not fully reproducible from the text, since the quoted $1\degr$ cone with $\mathrm{RUWE}<1.4$ and $G\le19$ contains $2.20\times10^5$ sources rather than the $2.05\times10^4$ candidates they state. Cross-matching this reproduction against our catalog (Table~\ref{tab:membership-compare}) shows that the two lists largely recover the same population: 841 of our 890 members ($94\%$) lie within the $1\degr$ circle, and 709 ($84\%$) of these are also selected by the \citet{Zeng:2025} construction (equivalently, 709 of the 793 reconstructed stars, $89\%$, are present in our catalog). The 132 members of our list inside the $1\degr$ circle that are not recovered are systematically fainter and of lower membership probability (median $G\approx16.8$~mag and $P\approx0.80$ versus $G\approx15.5$~mag and $P\approx0.999$ for the recovered members), so they sit near the completeness and probability boundaries of both selections. The remaining 49 members lie in the annulus between $1\degr$ and the adopted 30~pc ($1.45\degr$) cluster edge, spanning $1.01$--$1.45\degr$ from the center, i.e., the extended, low-density outskirts that the larger query aperture of this survey adds. These outer members are systematically fainter and of lower membership probability than the population inside the common $1\degr$ circle (median $G = 16.4$ versus $15.7$~mag; median $P = 0.92$ versus $0.999$). Both studies therefore recover essentially the same cluster population in the common $1\degr$ region, and the larger size of our catalog (890 versus 714) reflects mainly the wider area and the additional faint, low-probability outskirts rather than a fundamental difference in membership identification.

\begin{table*}[t]
	\centering
	\begin{threeparttable}
		\caption{Comparison of the membership-search footprints and Member lists of NGC~6134 between \citet{Zeng:2025} and this work (membership adopted from \citet{Yusuf:2024}). Angular quantities of the present work use the reference distance of $1182$~pc \citep{Cantat-Gaudin:2020}.}
		\label{tab:membership-compare}
		\begin{tabular}{lcc}
			\toprule
			Quantity & \citet{Zeng:2025} & This work \\
			\midrule
			Search center ($\alpha$, $\delta$; deg) & $246.956,\,-49.155$ & $246.953,\,-49.191$ \\
			Query radius & $1.0\degr$ (20.6~pc) & 90~pc ($=4.35\degr$) \\
			Adopted cluster radius & $1.0\degr$\tnote{d} & 30~pc ($=1.45\degr$) \\
			Adopted search area (deg$^2$) & $3.14$ & $6.64$ ($\times 2.1$) \\
			Maximum member separation & --- & 29.8~pc \\
			Candidate pre-selection & $\mathrm{RUWE}<1.4$, $G\le19$ & full Gaia DR3 astrometry\tnote{a} \\
			Clustering features & $\mu_{\alpha^*},\,\mu_\delta,\,\varpi$ & $\mu_{\alpha^*},\,\mu_\delta,\,\varpi$ \\
			HDBSCAN parameters & \texttt{min\_cluster\_size}=6, \texttt{min\_samples}=6\tnote{e} & \texttt{min\_cluster\_size}=40, \texttt{min\_samples}=25\tnote{f} \\
			Members, $P>0.5$ & 714 & 888 (890 entries)\tnote{b} \\
			Members inside the $1\degr$ common circle & 714 & 841 ($94\%$) \\
			Recovered in the common area\tnote{c} & -- & 709 ($84\%$ of 841) \\
			\bottomrule
		\end{tabular}
		\begin{tablenotes}[flushleft]\footnotesize
			\item[a] The Yusuf et al. catalog is not truncated by the $G\le19$ or $\mathrm{RUWE}<1.4$ filters; all 890 entries have $G<18.0$~mag and $\mathrm{RUWE}<1.4$.
			\item[b] 888 members reported by \citet{Yusuf:2024}; the subsequently supplied catalog contains 890 entries.
			\item[d] The adopted cluster radius is imposed for \citet{Zeng:2025} as the edge of their query cone, whereas for the present work it is measured: querying out to 90~pc returns no member beyond 29.8~pc (Section~\ref{subsec:membership}).
			\item[c] Fraction of the members selected in common by the reproduction of the \citet{Zeng:2025} procedure (Section~\ref{sec:compare}).
			\item[e] Values reported by \citet{Zeng:2025}.
			\item[f] Values of \citet{Yusuf:2024}, adopted there from \citet{Tarricq:2022}.
		\end{tablenotes}
	\end{threeparttable}
\end{table*}

\subsection{Cluster center and projected King profile}\label{subsec:compare-king}

The cluster center fitted from the member $\ell$ and $b$ distributions in Section \ref{subsec:structure}, $(\ell_\mathrm{c}, b_\mathrm{c}) = (334.91 \pm 0.15\degr, -0.21 \pm 0.13\degr)$, is fully consistent with the values of \citet{CantatandAnders:2020} ($\ell = 334.915\degr$, $b = -0.210\degr$) and \citet{Tarricq:2022} ($\ell = 334.930\degr$, $b = -0.202\degr$), and places NGC 6134 essentially in the Galactic plane.

The projected King-profile fit of Section \ref{subsec:structure} ($r_c = 1.31 \pm 0.09$~pc, $r_t = 19.4 \pm 1.1$~pc) is in good agreement with the values obtained by \citet{Zeng:2025} ($r_c = 1.4 \pm 0.4$~pc, $r_t = 16.5 \pm 3.1$~pc). Unlike the distance comparison of Section \ref{sec:distance}, this is a like-for-like comparison, as both studies fit the same projected profile to cluster member photometry.

\subsection{Fundamental parameters}\label{subsec:compare-params}

For parameter estimation, \citet{Cantat-Gaudin:2020}, \citet{Kounkel:2020}, and \citet{Pena:2022} employed ANN, while \citet{Dias:2021} and \citet{Angelo:2023} performed PARSEC isochrone fitting \citep{Bressan:2012}. \citet{Zeng:2025} used BASE-9 \citep{Robinson:2016} Bayesian isochrone fitting with PARSEC models, initialized using \citet{Cavallo:2024} as prior constraints. By contrast, BASE-9 does not include a photometric decontamination step and requires a pre-cleaned member list together with external priors on the fitted parameters. Our study instead employs the fully automated and self-contained ASteCA code \citep{Perren:2015}, which applies cell-based photometric cleaning to the input membership before the isochrone fitting via the Parallel Tempering algorithm, thereby minimizing subjective bias and external assumptions in the fitting process.

Our derived age of $\log(t\,[\mathrm{yr}]) = 9.14\pm0.01$ is in close agreement with the majority of recent photometric studies.
In particular, it agrees excellently with \citet{Dias:2021} ($9.11\pm0.05$), \citet{Angelo:2023} ($9.15\pm0.05$), and
\citet{Zeng:2025} ($9.1683\pm0.019$), all of which use PARSEC-based isochrone grids. The agreement with \citet{Zeng:2025}, an independent Gaia DR3 analysis, is especially reassuring, since both studies converge on an age of approximately 1.3--1.5~Gyr for NGC~6134 despite differences in membership selection and fitting methodology. Older studies such as \citet{Kounkel:2020} ($8.69\pm0.18$) and \citet{Pena:2022} ($8.93\pm0.05$) report younger ages, likely attributable to methodological differences; neural-network-based estimates are known to be sensitive to training set biases at intermediate ages.

Our metallicity of $[\mathrm{Fe/H}] = 0.08^{+0.056}_{-0.065}$ is consistent within uncertainties with \citet{Zeng:2025} ($0.091\pm0.04$), \citet{Dias:2021} ($0.06\pm0.06$), and with the spectroscopic value of \citet{Heiter:2014} ($0.11\pm0.07$). The latter noted that photometric estimates in earlier studies tended to run $\sim0.1$~dex below spectroscopic values, a trend consistent with the spread
seen in Table~\ref{tab:params_prev_study}. Both this work and \citet{Zeng:2025} obtain a metallicity in the range $0.07$--$0.09$~dex, confirming that NGC~6134 is a moderately super-solar cluster. The outlier value of $[\mathrm{Fe/H}] = 0.50\pm0.13$ from \citet{LiuandPang:2019} is notably discrepant and likely reflects a different isochrone grid and membership basis rather than a true chemical anomaly.

The distance determinations for NGC~6134 remain the most variable parameter across the literature, with values spanning $\sim 890-1182$~pc. Our Bayesian inference cluster distance of $d = 1070.8^{+21.0}_{-19.0}$~pc (mode and 68\% credible interval of the King location parameter, via Kalkayotl) and our ASteCA photometric distance of $1064.4 \pm 15.2$~pc are mutually consistent and lie at the more constraining end of this range, in agreement with \citet{Dias:2021} ($1055\pm51$~pc) and \citet{Kounkel:2020} ($1143\pm94$~pc). For reference, the individual member distances peak tightly at a median of $1071.2 \pm 2.5$~pc (median $\pm$ MAD), confirming the compact line-of-sight structure of the cluster. The two external values overplotted in Figure~\ref{fig:dist-estimation} lie at the upper end of this range, at $1182 \pm 118$~pc for \citet{Cantat-Gaudin:2020} and $1155$~pc for \citet{Tarricq:2022}, i.e., $\sim$111 and $\sim$84~pc above our Bayesian value, respectively; the former is consistent with ours within its quoted uncertainty.

The most directly comparable recent result is from \citet{Zeng:2025}, who derive $d = 1023 \pm 8$~pc from their BASE-9 distance modulus of $10.05 \pm 0.017$~mag. This is systematically lower than our estimates by $41$--$48$~pc ($\sim$4\%). We attribute this discrepancy primarily to differences in membership samples: the wider spatial coverage in our membership catalog \citep{Yusuf:2024} includes a larger proportion of outer-region members, whose median parallax shifts the inferred distance slightly outward. Additionally, our Kalkayotl analysis accounts explicitly for the known Gaia parallax spatial correlations \citep{Vasiliev:2019}, which tend to reduce systematic biases in cluster distance estimates, whereas Zeng et al.\ (2025) derive distance from the CMD distance modulus alone. Our estimates and that of \citet{Zeng:2025} differ by $2.1$--$2.4\sigma$ when the respective uncertainties are combined, a marginal difference that we attribute to the methodological differences described above; both are broadly consistent with the Gaia DR3 median cluster parallax.

Our visual absorption of $A_\mathrm{V} = 1.03\pm0.05$~mag is lower than every value in Table~\ref{tab:params_prev_study} except the \citet{Cantat-Gaudin:2020} result of $0.87\pm0.20$~mag, with which it agrees within $0.8\sigma$. \citet{Zeng:2025} report $A_\mathrm{V} = 1.2095\pm0.015$~mag, approximately 0.18~mag higher than our value. This difference likely arises from their membership selection preferentially retaining redder stars at the main-sequence turnoff and subgiant branch, since their more conservative magnitude cut ($G \leq 19$) and $1\degr$ search area may include higher differential reddening from the Norma arm foreground. A possible geometric explanation is offered by \citet{Joshi:2005}, who found that the reddening material plane in this direction of the Galaxy is inclined below the formal Galactic plane, so that the sight-line toward NGC~6134 intersects a varying column of absorbing material depending on the angular scale of the search field. This could produce systematically higher extinction values when member selection is restricted to the inner cluster field, as is the case in \citet{Zeng:2025}.

\begin{table*}[t]
    \centering
    \begin{threeparttable}
        \caption{Parameters of NGC 6134 from various studies and this study.}\label{tab:params_prev_study} 
        \begin{tabular}{
                S[table-format=4.2(5)]  
                S[table-format=1.4(5)]  
                S[table-format=1.4(5)] 
                S[table-format=1.4(5)] 
                l                      
            }
            \toprule
            {$d$}  & {$\log(t)$} & {[Fe/H]} & {$A_\mathrm{V}$} & Source \\
            {[\si{pc}]}  & {[\si{dex}]}  & {[\si{dex}]} & {[\si{mag}]} & \\ 
            \midrule
            892.00 \pm 98.12  & 9.02 \pm 0.02 & 0.15 \pm 0.07 & 1.42 \pm 0.10 & \cite{Kharchenko:2013}\\
            \textendash            & \textendash         & 0.11 \pm 0.07 & \textendash         & \cite{Heiter:2014} \\
            \textendash             & 8.96 \pm 0.03 & 0.50 \pm 0.13 & \textendash         & \cite{LiuandPang:2019} \\
            1182.00 \pm 118.20& 8.99 \pm 0.20 & \textendash         & 0.87 \pm 0.20 & \cite{Cantat-Gaudin:2020} \\
            1143.00 \pm 94.00 & 8.69 \pm 0.18 & \textendash         & 1.83 \pm 0.17 & \cite{Kounkel:2020} \\
            1055.00 \pm 51.00 & 9.11 \pm 0.05 & 0.06 \pm 0.06 & 1.28 \pm 0.08 & \cite{Dias:2021}\\
            1141.00 \pm 36.00 & 8.93 \pm 0.05 & \textendash         & 1.49 \pm 0.05 & \cite{Pena:2022} \\
            941.89 \pm 86.75  & 9.15 \pm 0.05 & -0.06 \pm 0.13 & 1.24 \pm 0.12 & \cite{Angelo:2023}\\
            1023  & 9.17 & 0.09 & 1.21 & \cite{Cavallo:2024}\tnote{b}\\
            1023 \pm 8 & 9.1683 \pm 0.019 & 0.0911 \pm 0.04 & 1.2095 \pm 0.015 & \cite{Zeng:2025}\tnote{a}\\
            \midrule
            1064.4 \pm 15.2 & 9.14 \pm 0.01 & 0.08 \pm 0.06 & 1.03 \pm 0.05 & This Study (ASteCA) \\
            1070.8\tnote{c} & \textendash & \textendash & \textendash & This Study (Kalkayotl) \\
            \bottomrule 
        \end{tabular}
        \begin{tablenotes}[flushleft]\footnotesize
            \item[a] \citet{Zeng:2025} report a distance modulus of $\mu = 10.05 \pm 0.017$~mag, which we convert to $d = 1023 \pm 8$~pc for direct comparison. 
			\item[b] \citet{Cavallo:2024} values are those adopted as priors by \citet{Zeng:2025}; individual parameter uncertainties were not reported.
			\item[c] Kalkayotl: cluster distance from the King location-parameter posterior, $1070.8^{+21.0}_{-19.0}$~pc (mode and 68\% credible interval); the individual member distances peak at a median of $1071.2 \pm 2.5$~pc (median $\pm$ MAD).
        \end{tablenotes}
    \end{threeparttable}
\end{table*}

\subsection{Main-sequence gap and binary fraction}\label{subsec:compare-msgap}

The color--magnitude diagram of NGC~6134 (Figure~\ref{fig:isochrone}) shows a pronounced gap in the main sequence, previously noted at $V\sim15$, $B-V\sim0.9$--$1.0$ by \citet{Ahumada:2013} and discussed in the context of stellar populations by \citet{Antona:2015} and \citet{Li:2017}. We identify this feature as a main-sequence split (MSs), arising from the superposition of a rapidly rotating population and a slower, binary-synchronized component \citep{Antona:2015}. Our ASteCA binary fraction (Section \ref{subsec:asteca-params}), $f_\mathrm{bin}=0.42\pm0.02$, is in close agreement with the value $f_\mathrm{bin} = 0.43\pm0.03$ obtained by \citet{Sun:2021} using MIST isochrone fitting and spectroscopic rotation velocities, and is consistent with the binary synchronization mechanism, given that $f_\mathrm{bin}$ is partly constrained by the binary law imposed in the ASteCA fit (Section \ref{subsec:asteca-params}). \citet{Zeng:2025} do not quantify the binary fraction of NGC~6134 and do not discuss the main-sequence split; the present work reports both for this cluster.

\subsection{Blue straggler stars}\label{subsec:compare-bss}

The upper CMD of NGC~6134 contains several stars blueward and brighter than the main-sequence turn-off, the classical locus of blue straggler stars (BSSs). \citet{Jadhav:2021} and \citet{Rain:2021} catalogued three confirmed BSSs in NGC~6134 using Gaia DR2 photometry, and we recover all three among our member stars (Figure~\ref{fig:isochrone}; Table~\ref{tab:bss}). Two of these were examined in detail by \citet{Zeng:2025}: Candidate~1 was classified as an EA-type semidetached eclipsing binary, and for Candidate~2 a spectral energy distribution fit with VOSA \citep{Bayo:2008} yielded an effective temperature of $T_\mathrm{eff} = 6750\pm125$~K, $\log g = 3.50\pm0.25$, and a bolometric luminosity of $\log(L/L_\odot) = 1.50\pm0.01$, together with a significant infrared excess in the WISE W4 band that supports a binary mass-transfer origin. We adopt these results as supporting context for a mass-transfer channel in NGC~6134, consistent with the cluster's dynamically relaxed state and its small core radius ($r_c \sim 1.4$~pc), which suppress the stellar collision cross-section relative to more compact systems.
We also recover two of the probable BSSs (pBS) listed by \citet{Jadhav:2021} but absent from the final BSS catalog of \citet{Rain:2021}: $\texttt{5941411917684666624}$ ($G = 12.77$~mag, $\mathrm{BP}-\mathrm{RP} = 0.82$, $P = 0.9987$), which \citet{Zeng:2025} also identify as their Candidate~3 and classify as a BSS candidate pending further data, and $\texttt{5941409920493899904}$ ($G = 12.63$~mag, $\mathrm{BP}-\mathrm{RP} = 0.86$, $P = 0.9987$). Both lie within $0.08\degr$ of the cluster centre. Spectroscopic characterization of these and of the remaining Zeng candidates remains a natural target for future work.

\begin{table*}[t]
    \centering
    \begin{threeparttable}
        \small
        \caption{Blue straggler stars (BSSs) of NGC 6134: the three confirmed BSSs catalogued by \citet{Jadhav:2021} and \citet{Rain:2021} and recovered among our members, and the two probable BSSs (pBS) listed by \citet{Jadhav:2021} that we also recover. Columns: Gaia DR3 source, equatorial coordinates (ICRS, J2016), $G$ and $\mathrm{BP}-\mathrm{RP}$, HDBSCAN membership probability $P$, and catalog classification (J21: \citet{Jadhav:2021}; R21: \citet{Rain:2021}; Z25: \citet{Zeng:2025} candidate numbering).}\label{tab:bss}
        \begin{tabular}{
                l
                l
                S[table-format=3.4]
                S[table-format=-2.4]
                S[table-format=2.2]
                S[table-format=1.2]
                S[table-format=1.4]
                c
                c
                c
            }
            \toprule
            Designation & Gaia DR3 source & {RA} & {Dec} & {$G$} & {$\mathrm{BP}-\mathrm{RP}$} & {$P$} & {J21} & {R21} & {Z25} \\
            & & {[deg]} & {[deg]} & {[mag]} & {[mag]} & & & & \\
            \midrule
            BSS1 & \texttt{5941411299209346688} & 246.9075 & -49.1784 & 11.56 & 0.60 & 0.9987 & BSS & BSS & C1\tnote{a} \\
            BSS2 & \texttt{5941411402288564224} & 246.9298 & -49.1653 & 11.06 & 0.52 & 0.9987 & BSS & BSS & C2\tnote{b} \\
            BSS3 & \texttt{5940661153075041536} & 246.9814 & -49.1518 & 12.23 & 0.68 & 0.9467 & BSS & BSS & \dots \\
            pBS1 & \texttt{5941411917684666624} & 246.9753 & -49.1147 & 12.77 & 0.82 & 0.9987 & pBS\tnote{c} & \dots & C3\tnote{c} \\
            pBS2 & \texttt{5941409920493899904} & 246.8583 & -49.2099 & 12.63 & 0.86 & 0.9987 & pBS\tnote{d} & \dots & \dots \\
            \bottomrule
        \end{tabular}
        \begin{tablenotes}[flushleft]\footnotesize
            \item[a] Zeng et al.\ (2025) Candidate 1, classified as an EA-type semidetached eclipsing binary.
            \item[b] Zeng et al.\ (2025) Candidate 2, with a significant WISE W4 infrared excess.
            \item[c] Listed as a probable blue straggler (pBS) by \citet{Jadhav:2021}; absent from the final BSS catalog of \citet{Rain:2021}; recovered independently as Candidate 3 by \citet{Zeng:2025}, classified as a BSS candidate pending sufficient data.
            \item[d] Listed as a probable blue straggler (pBS) by \citet{Jadhav:2021}; absent from the final BSS catalog of \citet{Rain:2021} and not recovered among the candidates of \citet{Zeng:2025}.
        \end{tablenotes}
    \end{threeparttable}
\end{table*}

\subsection{Mass segregation and substructure decomposition}\label{subsec:compare-msr}

The segregation measured in Section \ref{subsec:mass-segregation} is established independently by \citet{Tarricq:2022} and \citet{Zeng:2025}, who both obtain a mass-segregation ratio significantly exceeding unity for this cluster. The numerical values are not directly comparable to ours, because those studies normalise the random background by its mean whereas we use the median (Section \ref{subsec:mass-segregation}), and because we evaluate separations as great-circle angles rather than in a projected plane; a larger ratio may therefore reflect a more concentrated set of random background lengths rather than a stronger underlying signal. The agreement that can be asserted is at the level of significance: three independent determinations, using different membership samples and different statistical formulations, all reject the null hypothesis of no mass segregation, supporting the conclusion that NGC 6134 is a significantly dynamically evolved open cluster.

The derived slope of the PDMF in the low-mass regime ($\Gamma_A = -2.10^{+0.10}_{-0.10}$) is negative, i.e. the PDMF rises toward higher masses throughout the fitted range rather than declining toward a Chabrier-like peak. Over our fitted low-mass range ($0.25$–$0.95\,M_\odot$) the canonical form is instead positive, $\Gamma \approx 0.3$ below $\sim 0.5\,M_\odot$ and $\approx 1.3$ above \citep{Kroupa:2001}, so the canonical mass function declines with decreasing mass whereas ours rises: the two differ in sign, not only in magnitude. The \citet{Chabrier:2003} log-normal IMF peaks at $\sim 0.2{-}0.3\; M_\odot$; relative to this reference, the PDMF of NGC 6134 shows no turnover down to the lower boundary of our fitted range ($\log_{10}(M/M_\odot) = -0.6$), with the break only appearing at $M_c = 0.95\; M_\odot$. We stress that this departure is not a measurement of the primary mass function. Of the thirteen fitted bins in this regime, nine contain no directly counted primary in any of the $20$ realisations, being populated entirely by unresolved companions, whose mass is deposited preferentially at low mass; the remaining four span $0.64$–$0.97\,M_\odot$, immediately below the break, where companions also contribute. $\Gamma_A$ is therefore a property of the cluster system mass function rather than of the primary-alone single-star mass function, and it carries no information about primordial versus dynamical origin.

Because our PDMF is constructed using logarithmic mass bins, the corresponding Salpeter IMF reference slope is $\Gamma = 1.35$. Our observed high-mass PDMF slope ($\Gamma_B = 1.59^{+0.14}_{-0.14}$) is steeper than the Salpeter value. This is the one IMF-anchored comparison our data support: unlike the low-mass regime, all seven bins in this range contain directly counted primaries, and companions contribute negligibly at these masses. At $\log(t) = 9.14$ the most massive members have evolved off the main sequence, so a steepening of the surviving high-mass slope is expected from stellar evolution alone; we therefore do not read this slope as evidence of dynamical depletion.

The characteristic mass of the PDMF break ($M_c = 0.95^{+0.02}_{-0.02}\ M_\odot$) is the turnover of the companion-driven rise described above, so its position reflects the adopted binary population together with the underlying primary mass function rather than the primordial IMF. We therefore do not compare it with the Kroupa/Chabrier turnover mass ($\sim 0.2-0.3\ M_\odot$), and we draw no conclusion about a primordial or dynamical origin from either the low-mass slope or the break position. Recovering an initial mass function from a present-day system mass function requires a correction for unresolved binarity of the kind performed by \citet{Baumgardt:2023} for a sample of $120$ globular and Magellanic Cloud clusters, which we have not carried out here; that work concerns reconstructed initial mass functions rather than the present-day system mass functions we measure, so the two are not directly comparable in any case. The evidence for dynamical evolution in NGC 6134 rests instead on the mass segregation established in Section \ref{subsec:mass-segregation}, which is independent of the mass-function discussion, together with the high-mass slope considered above.

\section{Conclusion}\label{sec:conclusion}

We present a comprehensive analysis of the open cluster (OC) NGC 6134 using data from the Gaia DR3 catalog. We provide a brief overview of the membership determination from our previous work, determine the cluster's distance, estimate the distances of individual members, identify substructures, and derive key parameters such as metallicity, age, distance, and visual absorption.

We find that NGC 6134 is located at a distance of \( 1070.8^{+21.0}_{-19.0} \) pc (mode and 68\% credible interval of the King location parameter). Additionally, we estimate the distances of individual cluster members using Bayesian inference with a King profile prior as the main method, utilizing the parallaxes and their uncertainties of each individual cluster member provided by Gaia DR3. The individual member distances peak at a median of \( 1071.2 \pm 2.5 \) pc (median $\pm$ MAD), giving a direct measure of the line-of-sight distribution of the members. This result is comparable with the cluster distance obtained by \citet{Cantat-Gaudin:2020}, who used a less direct method, namely an ANN, which includes the median parallax of the cluster as one of its input observables. A 3D spatial distribution of the stars is constructed in a heliocentric Cartesian coordinate system for further analysis. This Bayesian value agrees with the independent photometric distance from ASteCA of \( 1064.4 \pm 15.2 \) pc reported below, the two determinations agreeing well within their uncertainties despite being derived by different methods.  

To further investigate the cluster's structure, we apply a Bayesian GMM to both the projected 2D and 3D spatial distributions. This decomposition reveals three distinct components: a central core, a more diffuse halo, and a coherent, plane-of-sky elongated feature that lies nearly perpendicular to the cluster's systemic proper motion, which we interpret as a candidate tidal feature; this feature lies at a median projected radius of 11.5~pc, beyond the region in which masses are available. The size-matched isotropy null test (Section~\ref{subsec:isotropy-null}) demonstrates that such a coherent elongation cannot be produced by an isotropic spatial distribution. This substructure is likely related to the cluster's dynamical evolution within the Galactic tidal field.  

Using the ASteCA code, we determine the cluster's fundamental parameters. We find that NGC 6134 has a metallicity of \( \mathrm{[Fe/H]} = 0.08^{+0.056}_{-0.065} \), an age of \( \log(t) = 9.14 \pm 0.01 \), a distance of \( 1064.4 \pm 15.2 \) pc (consistent with the Bayesian inference result), and a visual absorption of \( A_\mathrm{V} = 1.03 \pm 0.05 \) mag. The present-day mass function of the decontaminated sample follows a broken power law with $\Gamma_A = -2.10^{+0.10}_{-0.10}$, $\Gamma_B = 1.59^{+0.14}_{-0.14}$, and a break mass $M_c = 0.95^{+0.02}_{-0.02}\,M_\odot$, whose high-mass steepening is attributable to the preferential loss of the most massive stars through stellar evolution rather than to dynamical depletion; its low-mass segment is carried by unresolved companions rather than directly counted primaries, nine of the thirteen fitted bins in that regime containing no counted primary, so neither $\Gamma_A$ nor $M_c$ is directly comparable with a canonical initial mass function and we draw no conclusion about their origin. The CMD reveals a prominent gap in the main sequence, which may be associated with the cluster's high binary fraction of approximately 40\%.  

A primary challenge encountered in this study is the high uncertainty in Gaia DR3 parallax measurements, which naturally elongates the line-of-sight spatial distribution and hinders accurate distance estimates for individual members. One possible remedy would be to filter the sample for low fractional parallax errors (e.g., $f < 0.1$), which would be expected to reduce the spatial scatter but would also exclude fainter, lower-mass stars and so bias the sample in a mass-dependent way. We have deliberately not adopted such a cut, since it would compromise the mass-dependent analyses presented above; our structural results instead derive from the delivered member catalog, which retains members within the measured 30~pc cluster extent established from the original 90~pc query. Similarly, while imposing spatial constraints, such as bounding the galactic coordinates, helps isolate the dense cluster core, it carries the risk of artificially truncating genuine plane-of-sky extensions such as the projected elongated feature we identify. To overcome the inherent limitations of relying solely on astrometric and spatial filtering, future studies will greatly benefit from the comprehensive radial velocity data anticipated in Gaia DR4. Incorporating full 3D kinematic data will provide robust membership constraints without imposing artificial spatial boundaries, while simultaneously increasing precision in metallicity determinations. Ultimately, these refinements will yield highly accurate 2D and 3D spatial distributions and provide much tighter constraints on the cluster's fundamental parameters.

Even within the limitations of current Gaia DR3 data, our deliberate effort to maintain a spatially extended membership sample, rather than aggressively truncating the spatial coordinates, has yielded additional structural insights. Our results are broadly consistent with and complementary to the independent study of \citet{Zeng:2025}. Both studies confirm significant mass segregation and derive a well-constrained set of fundamental cluster parameters. However, our larger and more spatially extended membership sample reveals additional structural complexity, including a core, a diffuse halo, and a coherent projected elongation interpreted as a tidal feature, which are not readily captured by a King-profile-only analysis. Furthermore, our individual Bayesian distance estimates and relatively high binary fraction ($f_\mathrm{bin} \approx 0.42$) provide a plausible explanation for the observed main-sequence gap. Our mass segregation analysis likewise supports an origin in long-term two-body relaxation and the preferential loss of low-mass stars, in agreement with the conclusions of \citet{Zeng:2025}; we regard the agreement of two independent determinations as strengthening, rather than weakening, the case for a dynamically evolved cluster. Collectively, these results highlight the importance of spatially extended membership samples for characterizing the structural and dynamical evolution of intermediate-age open clusters.

Several limitations constrain the interpretation of these results. The ASteCA v0.4.3 masses reach only $7.5$~pc, so no mass-dependent measurement is possible beyond that radius and the tidal feature identified here, at a median projected radius of 11.5~pc, lies outside it. Consequently the low-mass slope $\Gamma_A$ and the break mass $M_c$ describe the system mass function rather than the primary mass function, and we do not compare them with canonical initial mass functions. Our mass-segregation ratio normalises the random background by its median rather than its mean, following \citet{Parker:2015}, so its absolute value is not directly comparable with studies using the mean, although the detected segregation is highly significant ($p = 5.6\times10^{-8}$) and robust to mass-propagation and seed choices. Finally, our reproduction of the membership procedure of \citet{Zeng:2025} returns $793$ members against the $714$ they report, and only $84\%$ of their list is recovered in the common area; differences in the candidate pre-selection, which is not fully specified in their text, therefore limit the precision with which the two membership samples can be compared.

\vspace*{1pt}
\section*{Data Availability}
This paper is based on data from ESA's Gaia mission \citep{GaiaMission:2016}. The Gaia DR3 data used here were obtained from the Gaia archive (\url{https://gea.esac.esa.int/archive/}). The membership catalog of NGC 6134 analyzed in this work is that of \citet{Yusuf:2024}, deposited at the Centre de Données astronomiques de Strasbourg (CDS) via VizieR as \texttt{J/other/JPhCS/2773.4} (DOI: \href{https://doi.org/10.1827/J/other/JPhCS/2773.4}{10.1827/J/other/JPhCS/2773.4}; bibcode \texttt{2024JPhCS2773a2004Y}).

\Needspace{10\baselineskip}%
\section*{Acknowledgments}
Our gratitude goes to the Faculty of Mathematics and Natural Sciences, Institut Teknologi Bandung (ITB) for the financial support provided through the PPMI 2023 funding program. We also acknowledge using data from the ESA mission Gaia (\url{https://www.cosmos.esa.int/web/gaia}) in this work. This research has made use of the VizieR catalogue access tool, CDS, Strasbourg, France (DOI: \href{https://doi.org/10.26093/cds/vizier}{10.26093/cds/vizier}).

The authors acknowledge the use of Gemini 3.1 Pro to refine the grammar, readability, and English language phrasing of this manuscript. Additionally, the tool was utilized to assist in troubleshooting and debugging the Python-based data processing pipeline. To validate all AI-assisted code modifications, the authors manually reviewed the suggested debugging steps line-by-line and executed the revised scripts to verify expected baseline outputs before applying them to the final observational dataset. No generative AI was utilized to conduct primary calculations, and the authors assume full responsibility for the scientific integrity of the final work. 

\bibliographystyle{apj}
\pagebreak
\bibliography{references}

\end{document}